\pdfoutput=1
\documentclass[pdflatex,sn-mathphys-ay]{sn-jnl}

\usepackage{graphicx}%
\usepackage{multirow}%
\usepackage{amsmath,amssymb,amsfonts}%
\numberwithin{equation}{section}
\usepackage{amsthm}%
\usepackage{mathrsfs}%
\usepackage[title]{appendix}%
\usepackage{xcolor}%
\usepackage{textcomp}%
\usepackage{manyfoot}%
\usepackage{booktabs}%
\usepackage{algorithm}%
\usepackage{algorithmicx}%
\usepackage{algpseudocode}%
\usepackage{listings}%
\usepackage{bm}
\usepackage{newtxtext}
\usepackage{natbib}
\usepackage{setspace}
\usepackage{ulem} 
\usepackage{accents} 
\usepackage{subfigure}
\theoremstyle{thmstyleone}%
\theoremstyle{thmstyletwo}%

\theoremstyle{thmstylethree}%

\begin{document}

\title[Article Title]{Heat vortices and the vorticity triplet in kinetic phonon transport: a fluid-dynamical perspective}


\author*[1]{\fnm{Tao} \sur{Chen}}\email{chentao2023@njust.edu.cn}
\author*[2,3]{\fnm{Chuang} \sur{Zhang}}\email{zhangc26@zju.edu.cn}

\affil*[1]{\orgdiv{School of Physics}, \orgname{Nanjing University of Science and Technology}, \orgaddress{ \city{Nanjing}, \postcode{210094}, \state{Jiangsu}, \country{China}}}
\affil*[2]{\orgdiv{Center for Digital-Physics Interdisciplinary Study}, \orgname{Zhejiang University}, \orgaddress{ \city{Hangzhou}, \postcode{310058}, \state{Zhejiang}, \country{China}}}
\affil*[3]{\orgdiv{Institute of Fundamental and Transdisciplinary Research}, \orgname{Zhejiang University}, \orgaddress{ \city{Hangzhou}, \postcode{310058}, \state{Zhejiang}, \country{China}}}
\author[4]{\fnm{Tianshu} \sur{Liu}}
\affil[4]{\orgdiv{Department of Mechanical and Aerospace Engineering}, \orgname{Western Michigan University}, \orgaddress{\city{Kalamazoo}, \postcode{49008}, \state{Michigan}, \country{USA}}}


\abstract{From a novel fluid-dynamical perspective, the present study investigates the heat-current vortices and heat vorticity in heat conduction process of solid materials. The heat vorticity $\bm{\omega}$, defined as the curl of the heat-flux field, is decomposed into the orbital-rotation $\bm{R}$ and spin $\bm{S}$, which collectively constitute the heat-vorticity triplet $(\bm{R},\bm{S},\bm{\omega})$, extending beyond the existing single-vorticity research paradigm. As an illustrative example, kinetic phonon transport in a 2D porous solid structure is examined based on the numerical solutions of the Peierls-Boltzmann equation under the Callway dual-relaxation-time approximation. The physical characteristics of heat vortices and the vorticity triplet patterns are analyzed across different kinetic regimes in the phase diagram of the normal (N) and resistive (R) scattering processes. Although heat vortices can emerge in both the ballistic and hydrodynamic regimes, the distributions of the vorticity triplet exhibit pronounced differences, which in turn serve as effective signatures for distinct kinetic regimes. In the hydrodynamic limit, the boundary generation mechanism of heat vorticity is further elucidated based on the Guyer-Krumhansl (G-K) equation generated from the truncated Chapman-Enskog (C-E) expansion. The boundary heat vorticity flux (BHVF) is proposed as a new concept for describing the local rate at which heat vorticity is generated at the boundary of a solid material. The general expression for the BHVF is rigorously derived for phonon hydrodynamics, and for the 2D steady hydrodynamic phonon transport case, we find that the BHVF is solely attributed to the tangential component of the net heat flux deviating from the classical Fourier heat flux. Furthermore, we show that the total vorticity contained in a control volume arises from the spatiotemporal accumulation of the BHVF, weighted by the memory kernel arising from the R scattering. The proposed theory and
findings herein could be valuable for understanding the diverse vortical-flow phenomena in condensed matter systems.}




\maketitle
\tableofcontents
\section{Introduction}\label{Introduction}
Heat conduction is a long-standing discipline whose rational foundations trace back to the seminal work of Joseph Fourier, who formulated the constitutive law that now bears his name in his landmark treatise~\citep{Fourier1822theorie}. Precise characterization and in-depth understanding of heat transport are fundamentally essential not only to rational thermodynamics and kinetic theory~\citep{Boltzmann1964lectures,Gibbs1902elementary,Muller1998,Guo2015Phonon}, but also a broad spectrum of engineering applications such as nanoscale heat transport in semiconductors~\citep{Bird1960transport,Deen2011analysis}. The physical mechanisms of heat transfer can be interpreted via multiple heat carriers, including molecules, electrons, photons, and phonons~\citep{Kaviany2008heat}. Among them, the concept of the phonon was originally introduced by~\citet{Peierls1929zur} to describe heat conduction in dielectric solids. Phonons represent quantized lattice vibrations that obey the Hamiltonian structure of kinetic theory~\citep{Grmela2011} and are responsible for energy transport in crystalline materials~\citep{Sykora2023}.

Since phonons can be generated or annihilated during scattering events, their number is generally not conserved. The resulting phonon transport behavior falls into distinct kinetic regimes, depending on the relative strengths of resistive (R) and normal (N) scattering processes~\citep{Hardy1970phonon,zhang2021heat}.
Both R and N scattering conserve energy whereas only N scattering preserves crystal momentum~\citep{Guo2015Phonon,Shang2020HeatVortex}.
At room temperature, heat conduction in bulk solid materials is well described by the classical Fourier's law, corresponding to the diffusive regime dominated by R scattering~\citep{Chen2005nanoscale}. However, ballistic and hydrodynamic phonon-transport regimes emerge upon reducing the system size~\citep{Hardy1970phonon,Majumdar1993}. In the ballistic regime, where both R and N scattering are very weak, the boundary-blockage effect dominates phonon transport, leading to a significant reduction in thermal conductivity with decreasing system characteristic length~\citep{Majumdar1993}. At extremely low temperatures or in low-dimensional materials, the hydrodynamic regime can emerge when N scattering greatly exceeds Umklapp (U) scattering or R scattering~\citep{Hardy1970phonon}. 
Derived from the linearized Chapman-Enskog multiscale expansion~\citep{Chapman1953,Callaway1959}, phonon hydrodynamics can be described by the macroscopic Guyer-Krumhansl (GK) equation for the heat flux $\bm{q}$ with appropriate boundary conditions, which resembles the Navier-Stokes (NS) equation in fluid dynamics at low Reynolds numbers, albeit without the nonlinear convection term $\bm{q}\bm{\cdot}\bm{\nabla}\bm{q}$~\citep{Guyer1966s,Guyer1966t,Shang2020HeatVortex,zhang2021heat}.
However, this nonlinear term emerges in a novel kinetic theory of phonon hydrodynamics developed by~\citet{Sykora2023}. This theory was rigorously derived via Poisson-bracket reduction without adopting linearization for reversible evolution. By virtue of the Chapman-Enskog expansion, the zeroth-order approximation generalizes the Maxwell-Cattaneo-Vernotte (MCV) equation in the one-dimensional (1D) case~\citep{Cattaneo1948} and leads to the Fourier-Crocoo equation, similar to the Crocco equation in aerodynamics~\citep{Anderson2017fundamentals}, while the first-order result generates the viscous diffusion term, as in the GK equation~\citep{Guyer1966s,Guyer1966t}. We note that the convection term also appears in the earlier works of~\citet{Banach2008chapman} and~\citet{Simoncelli2020}. Characteristic hydrodynamic phonon phenomena include second-sound propagation~\citep{Joseph1989heat,Huberman2019}, phonon Poiseuille flow~\citep{Machida2018observation,Li2018role}, and Knudsen minimum~\citep{Ding2018phonon,LiX2019PRB}, all of which have been studied both theoretically and experimentally.

Recently, within the framework of the Boltzmann-BGK phonon kinetic theory and viscous heat equations (VHEs)~\citep{Bhatnagar1954,Callaway1959,Guyer1966s}, heat (current) vortices with closed heat-flux streamlines have been observed as a novel phonon transport phenomenon emerging in the ballistic, hydrodynamic, and other transition regimes~\citep{Zhang2019implicit,Shang2020HeatVortex,zhang2021heat,Rayamendoza2022,DiLucente2026vortices}. Phonon hydrodynamics also exhibits a fascinating von K\'{a}rm\'{a}n vortex street at a relatively high heat-Reynolds number within a more general kinetic theory of phonons~\citep{Sykora2023}. However,~\citet{Huberman2018} predicted that the Reynolds number of graphene at room temperature is around one, suggesting that high Reynolds numbers may be difficult to realize in practical materials. To date, theoretical predictions of phonon turbulence are still subject to controversy and uncertainties, and no experimental observations have been reported~\citep{Zhang2025review}. These studies point to intriguing analogies between phonon transport in solids and fluid dynamics.

In fluid mechanics, vorticity is defined as the curl of the velocity field (i.e., $\bm{\omega}_{f}\equiv\bm{\nabla}\times\bm{u}_{f}$, where $\bm{u}_{f}$ denotes the fluid velocity), which has classically been interpreted as twice the angular velocity of a fluid
volume element in the classical Cauchy-Stokes theorem~\citep{Truesdell1954,Batchelor1967}.
Vortices are the sinews and muscles of fluid motion~\citep{Kuchemann1965}. From
a phenomenological morphological perspective, the basic forms of vortex structures at high Reynolds numbers typically manifest as columnar/axial vortices and sheet-like shear layers. The rolling-up of thin shear layers is the only known mechanism for the rapid formation of axial vortices~\citep{Chen2026General}. However, unlike vorticity, a precise and universal definition of a vortex remains an open issue under ongoing debate~\citep{Lugt1979,Liu2018Rortex}. For the analysis of coherent vortical structures, the vorticity representation offers superior physical intuition relative to the velocity field~\citep{Wu2015VorticalFlows,YangYue2020,Chen2023Lamb}. However, vorticity alone could not distinguish pure shearing motions from
true swirling motions. Recent theoretical studies have unraveled that vorticity can be decomposed into two elementary modes: the rigid-rotation mode $\bm{R}_{f}$ and the spin/shear mode $\bm{S}_{f}$, adopting either the characteristic algebraic approaches~\citep{Kolar2007vortex,Liu2018Rortex,WangYQ2019,LiuCQ2025} or material/field methods on the basis of oriented line and surface elements~\citep{Chen2025Kinematic,Chen2026General}. This reveals that axial vortices are dominated by the rigid-rotation mode, whereas shear layers are dominated by the spin/shear mode. These three quantities constitute the vorticity triplet $(\bm{R}_{f},\bm{S}_{f},\bm{\omega}_{f})$ for general vortical flows.~\citet{Chen2026Kinematic} proposed an operator-form vorticity decomposition, which provides a complete rational description for the fundamental forms of vortical structures with different spatial orientations. Beginning with the NS equations, the Lagrangian evolution equations for the rigid-rotation and spin modes were derived within the characteristic algebraic frame, which can be applied to study the dynamical interaction and mutual transformation between axial vortices and shear layers~\citep{Chen2026Lagrangian}.

By analogy with the concept of vorticity in fluid dynamics, heat/thermal vorticity is naturally defined as the curl of the heat flux field $(\bm{\omega}\equiv\bm{\nabla}\times\bm{q})$~\citep{zhang2021heat,Sykora2023,TURPRATS2024125464}. Clearly, when the thermal conductivity coefficient is a constant, the heat vorticity vanishes in the diffusive regime governed by Fourier's law, due to the alignment between the heat flux and the temperature gradient~\citep{zhang2021heat}. The appearance of non-null vorticity in other transport regimes demonstrates the significant deviations from diffusive transport~\citep{TURPRATS2024125464}. The phonon hydrodynamics arising from thermal compressibility and vorticity are separated by combining the viscous heat equations (VHEs) and the Helmholtz decomposition~\citep{DiLucente2026vortices,Dragasevic2026}. In addition, vorticity serves as the macroscopic signature of viscous electron flows~\citep{Levin2018PRB,Danz2020}, predicting that vorticity generated by viscous shear flow induces backflow against the applied electric field and yields negative nonlocal voltage~\citep{Levitov2016}. To the authors' knowledge, the published studies of heat vortices are rather limited  compared to the century-long history of fluid turbulence. Further theoretical analyses on heat vortices, vorticity, and its generation mechanisms remain to be explored.

The present article is structured as follows. In Section \ref{PB}, we introduce the Peierls-Boltzmann equation for phonon transport under the Callaway approximation, which serves as the theoretical foundation of this study. Section~\ref{Kinetic_regimes} discusses the phase diagram of the kinetic regimes of phonon transport. In Section~\ref{Heat vortices and vorticity decomposition}, we generalize the recently proposed vorticity decomposition method in fluid mechanics to the orbital-spin decomposition of heat vorticity based on the macroscopic phonon drift velocity. Section~\ref{Numerical simulations and analyses} presents the numerical results and analyzes the heat vortices, the vorticity triplet, and the mechanisms boundary heat vorticity generation in solids. Finally, conclusions and outlook are provided in Section~\ref{Conclusions and Outlook}.

\section{Peierls-Boltzmann Equation for Phonon Transport under the Callaway Approximation}\label{PB}
The phonon Boltzmann equation (BTE), incorporating the Callaway dual-relaxation-time approximation for the collision term which describes phonon scattering, is expressed as~\citep{Peierls1929zur,Callaway1959}
\begin{eqnarray}\label{BTE1}
	\frac{\partial f}{\partial t}+\bm{v}\bm{\cdot}\bm{\nabla}f=\frac{f_{R}^{eq}-f}{\tau_R}+\frac{f_{N}^{eq}-f}{\tau_N},
\end{eqnarray}
where $f=f(\bm{x},p,\omega,\bm{s},t)=f(\bm{x},p,\bm{k},t)$ denotes the phonon distribution function; $\bm{x}$ is the spatial location and $t$ is time; $p$ is the phonon polarization; $\bm{k}$ is the 2D isotropic wave vector, with $\bm{s}\equiv\bm{k}/\lVert\bm{k}\rVert$ indicating its orientation. Under the Debye and gray approximations~\citep{Debye1912,Chen2005nanoscale}, the group velocity is determined from the dispersion relation as $\bm{v}\equiv\bm{\nabla}_{\bm k}\omega=v_{g}\bm{s}$, where $\omega=\omega(\bm{k})$ is the phonon angular frequency and $v_{g}=\lVert\bm{v}\rVert$ is the group velocity magnitude. $\tau_R$ is the effective relaxation time of resistive (R) scattering processes that destroy momentum conservation, including umklapp (U) scattering, impurity scattering, isotope scattering, etc. $\tau_{N}$ is the effective relaxation time of normal (N) scattering  processes which conserve both momentum and energy.
For the R scattering, the energy conservation drives the system towards the equilibrium Bose-Einstein distribution, i.e.,~\citep{Bose1924,Einstein1924}
\begin{equation}\label{equilibrium}
f_{R}^{eq}(T)=\left[\exp\left(\frac{\hbar\omega}{k_B T}\right)-1\right]^{-1},
\end{equation}
where $k_{B}$ is the Boltzmann constant, $\hbar\equiv h/2\pi$ is reduced Planck constant, and $T$ is the temperature. The N scattering preserves both momentum and energy, which relaxes the distribution function toward a displaced Bose-Einstein equilibrium as~\citep{Guyer1966s}
\begin{equation}\label{displace_equilibrium}
	f_{N}^{eq}(T,\bm{u})=\left[\exp\left(\frac{\hbar\omega-\hbar\bm{k}\bm{\cdot}\bm{u}}{k_B T}\right)-1\right]^{-1}.
\end{equation}
Here, $\bm{u}$ is referred to as the drift velocity (constant at a given location and shared by all phonon modes)~\citep{Guo2015Phonon,Yang2019}. In short, phonon
Boltzmann equation~\eqref{BTE1} describes a balance among the variation, advection, and scattering processes of phonons.

In practical calculation, by introducing the deviational energy density distribution function, defined as the deviation from a reference equilibrium,
\begin{eqnarray}\label{transform}
e=\frac{\hbar\omega D}{2\pi}\left(f-f_{R}^{eq}(T_{0})\right),~~D\equiv\frac{\lVert\bm{k}\rVert}{2\pi\lVert\bm{v}\rVert},
\end{eqnarray}
we rewrite~\eqref{BTE1} in the form of~\citep{GUOY2017heat}
\begin{subequations}
\begin{equation}\label{BTE2}
	\frac{\partial e}{\partial t}+\bm{v}\bm{\cdot}\bm{\nabla}e=\frac{e_{R}^{eq}-e}{\tau_R}+\frac{e_{N}^{eq}-e}{\tau_N},
\end{equation}
where the equilibrium distribution functions in~\eqref{equilibrium} and~\eqref{displace_equilibrium} are transformed into
\begin{equation}\label{equi_R}
	e_{R}^{eq}(T)=\frac{\hbar\omega D}{2\pi}\left(f_{R}^{eq}(T)-f_{R}^{eq}(T_{0})\right),
\end{equation}
\begin{equation}\label{equi_N}
	e_{N}^{eq}(T,\bm{u})=\frac{\hbar\omega D}{2\pi}\left(f_{N}^{eq}(T,\bm{u})-f_{R}^{eq}(T_{0})\right).
\end{equation}	
\end{subequations}
Assuming a small temperature deviation from the reference temperature $T_{0}$ (i.e., $\Delta T=\lvert T-T_0\rvert\ll 1$) and a small drift velocity (i.e., $\bm{k}\bm{\cdot}\bm{u}\ll \omega$),~\eqref{equi_R} and~\eqref{equi_N} can be linearized as 
\begin{subequations}
	\begin{gather}
		e_{R}^{eq}(T)\approx \frac{C}{2\pi}\left(T-T_0\right), \\
		e_{N}^{eq}(T)\approx \frac{C}{2\pi}\left(T-T_0\right)+CT\frac{\bm{k}\bm{\cdot}\bm{u}}{2\pi\omega},
	\end{gather}
\text{where the mode specific heat at $T_{0}$ is evaluated as}
	\begin{equation}
		C(\omega,p,T_0)=2\pi\left.\frac{\partial e_{R}^{eq}}{\partial T}\right\rvert_{T=T_0}.
	\end{equation}
\end{subequations}

Once the energy density distribution function $e$ is obtained, the local temperature $T$ and the heat flux $\bm{q}$ are calculated as, respectively,
\begin{subequations}
	\begin{equation}
		T=T_{0}+\left(\sum\limits_{p}\int Cd\omega\right)^{-1}\left(\sum\limits_{p}\int\int ed\Omega d\omega\right),
	\end{equation}
	\begin{equation}
		\bm{q}=\sum\limits_{p}\int\int\bm{v}\bm{e}d\Omega d\omega.
	\end{equation}
\end{subequations}
Here, $d\Omega$ and $d{\omega}$ are the differential elements for the 2D solid angle and frequency space, respectively. It is noted that the temperature $T$ is not the true thermodynamic temperature, which serves as a measure of the local energy density~\citep{Chen2005nanoscale}. The following constraints can be derived from the conservation properties for the $R$ and $N$ scatterings:
\begin{subequations}
	\begin{align}
		0 &= \sum_{p} \iint \frac{e_R^{\text{eq}}(T_R) - e}{\tau_R} \, d\Omega d\omega, \\
		0 &= \sum_{p} \iint \frac{e_N^{\text{eq}}(T_N) - e}{\tau_N} \, d\Omega d\omega, \\
		0 &= \sum_{p} \iint \frac{\boldsymbol{k}}{\omega} \frac{e_N^{\text{eq}}(T_N,\boldsymbol{u}) - e}{\tau_N} \, d\Omega d\omega,
	\end{align}
\end{subequations}
where $T_{R}$ and $T_{N}$ are the local pseudo-temperatures to enforce the conservation laws~\citep{Guo2015Phonon,GUOY2017heat}. Consequently, we obtain
\begin{subequations}
	\begin{align}
		T_R &= T_0 +  \left( \sum_p \int \frac{C}{\tau_R} \, d\omega \right)^{-1}\left( \sum_p \int \frac{\int e \, d\Omega}{\tau_R} \, d\omega \right), \\
		T_N &= T_0 +  \left( \sum_p \int \frac{C}{\tau_N} \, d\omega \right)^{-1}\left( \sum_p \int \frac{\int e \, d\Omega}{\tau_N} \, d\omega \right), \\
		\bm{u} &= \frac{2}{T_N}\left(\sum\limits_{p}\int \frac{|\bm{k}|^2}{\omega^2} \frac{C}{\tau_N} \, d\omega\right)^{-1} \sum\limits_{p} \int \int \frac{\bm{k}}{\omega} \frac{e}{\tau_N} \, d\Omega \, d\omega.
	\end{align}
\end{subequations}

Note that the zeroth and first-order moments of~\eqref{BTE2} yields the macroscopic equations~\citep{Guyer1966s,Guyer1966t,Guo2015Phonon,Shang2020HeatVortex,zhang2021heat}
\begin{subequations}\label{eq29ab}
	\begin{align}
		\frac{\partial E}{\partial t} + \bm{\nabla}\bm{\cdot}\bm{q} &= 0, \label{eq29a} \\
		\frac{\partial\bm{q}}{\partial t} + \bm{\nabla}\bm{\cdot}\bm{Q} &= -\frac{\bm{q}}{\tau_{R}}, \label{eq29b}
	\end{align}
\end{subequations}
where the local deviational energy and the heat-flux current tensor are, respectively, given by
\begin{subequations}
	\begin{align}
		E &= \sum\limits_{p}\int\int\bm{e}d\Omega d\omega, \\
		\bm{Q} &= \sum\limits_{p}\int\int\bm{vv}{e}d\Omega d\omega.
	\end{align}
\end{subequations}

\section{Kinetic Regimes of Phonon Transport}\label{Kinetic_regimes}
\begin{figure}[t]
	\centering
	\includegraphics[width=0.9\columnwidth,trim={4cm 5cm 4cm 3.5cm},clip]{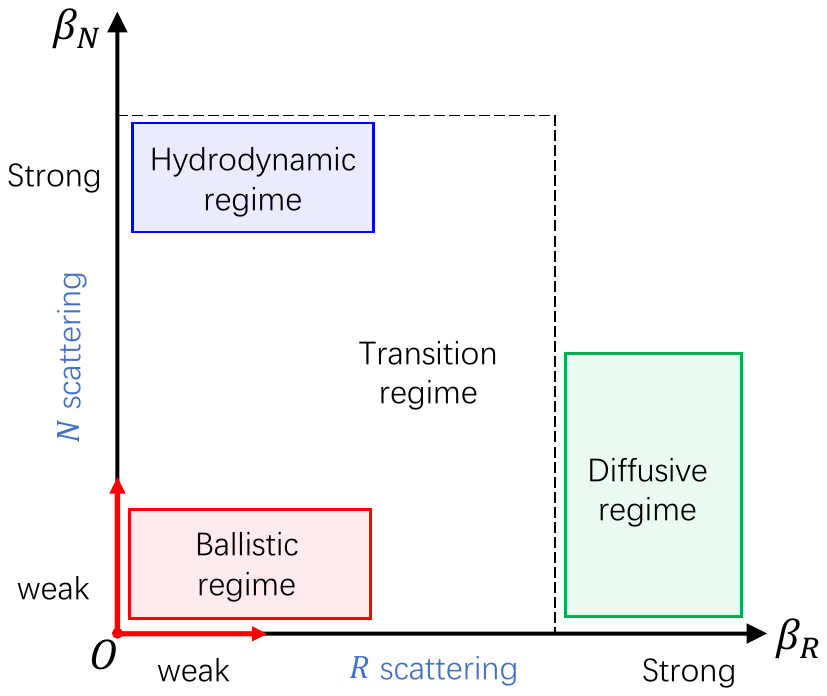}
	\centering
	\caption{Phase diagram of kinetic regimes for phonon transport classified by the inverse Knudsen-number pair $(\beta_{R},\beta_{N})$.} 
	\label{regimes}
\end{figure}

In the present study, we choose the specific heat $C$, the temperature $T_0$, the system characteristic length $L$, and the group velocity magnitude $v_{g}$ as the reference quantities for normalization.
Consequently, the kinetic system is governed by a pair of Knudsen numbers $({\rm Kn}_{R},{\rm Kn}_{N})$, whose inverses $({\beta}_{R},{\beta}_{N})$ are given by
\begin{eqnarray}
\beta_{R}\equiv{\rm Kn}_{R}^{-1}=\frac{L}{v_g \tau_{R}},~~\beta_{N}\equiv{\rm Kn}_{N}^{-1}=\frac{L}{v_g \tau_{N}},
\end{eqnarray}
which indicates the strength of $R$ and $N$ scatterings, respectively. The phonon mean free paths of the $R$ and $N$ processes are $\lambda_{R}=v_{g}\tau_{R}$ and $\lambda_{N}=v_{g}\tau_{N}$, respectively. Figure~\ref{regimes} shows four distinct kinetic regimes of phonon transport in terms of the parameter pair $({\beta}_{R},{\beta}_{N})$. The limiting case with $\beta_{R}\ll{1}\ll\beta_{N}$ corresponds to the hydrodynamic regime, while the limiting case with $\beta_{R}\ll{1}$ and $\beta_{N}\ll{1}$ corresponds to the ballistic regime. The limiting case with $\beta_{R}\gg{1}$ corresponds to the diffusive regime. Based on the numerical solutions of the phonon BTE for nanoscale porous silicon structures arranged in periodic arrays,~\citet{zhang2021heat} observed the emergence of heat current vortices (i.e., closed heat-flux streamlines) in the hydrodynamic, ballistic, and other transition regimes. However, further theoretical analyses of features and formation mechanisms of heat vortices from a fluid-dynamical perspective were not achieved therein.

\section{Heat Vortices and Vorticity Decomposition}\label{Heat vortices and vorticity decomposition}
\subsection{Heat-Flux Gradient Tensor and Heat Vorticity}
The classical framework of fluid mechanics (more specifically, vorticity and vortex dynamics) is based on a single vorticity measure for rotational flows~\citep{Batchelor1967,Wu2015VorticalFlows}. Vorticity equals twice the angular velocity of a fluid volume element, yet the complete rotational kinematics of oriented line and surface elements, which possess additional degrees of freedom, have been largely overlooked until recent work~\citep{Chen2026General}.
Recent theoretical studies have revealed two elementary modes hidden within vorticity from different perspectives, namely the rigid-rotation mode and the spin mode. By combining field-theoretic analysis with differential geometry,~\citet{Chen2025Kinematic} discussed the rotational kinematics of a pair of orthogonal line elements across all possible scenarios, and proposed a direction-dependent vorticity decomposition (DVD) based on local velocity orientation for generic 2D flows. Subsequently, a general theory of fluid-element rotation and intrinsic vorticity decomposition was reported in~\cite{Chen2026General}. More recently, a unifying principle proposed by~\citet{Chen2026Kinematic}
links these distinct vorticity modes to two communicative vorticity/vortex operators within a novel operator-form vorticity decomposition framework. We note that the application of vorticity decomposition is not limited to fluids, and an extended theory for general phonon transport in 2D solid materials is proposed below.

In analogy with the conventional fluid mechanics, the heat-flux gradient tensor admits the following symmetric-antisymmetric decomposition (SAD):
\begin{subequations}
	\begin{gather}
		\bm{A}\equiv\bm{\nabla}\bm{q}=\bm{D}+\bm{\varOmega}, \\
		\bm{D}\equiv\frac{1}{2}\left(\bm{\nabla}\bm{q}+\bm{\nabla}\bm{q}^{\rm T}\right),~~	\bm{\varOmega}\equiv\frac{1}{2}\left(\bm{\nabla}\bm{q}-\bm{\nabla}\bm{q}^{\rm T}\right),
	\end{gather}
\end{subequations}
where $\bm{D}$ denotes the heat-flux strain-rate tensor, and $\bm{\varOmega}$ the heat-flux rotation-rate tensor. 
The dual vector of $\bm{\varOmega}$ defines \textit{the heat vorticity} (or the thermal vorticity) as the curl of $\bm{q}$:
\begin{equation}
\bm{\omega}\equiv\bm{\nabla}\times\bm{q}.
\end{equation}
Note that the heat flux $\bm{q}$ is shown to be proportional to the drift velocity $\bm{u}$ whereas different degrees of approximation to the equilibrium state lead to discrepancies in the proportionality coefficients~\citep{Zhang2021Transient,Guo2015Phonon}. For example, the sub-first-order approximation of the zeroth-order Chapman-Enskog expansion of the phonon Boltzmann equation yields $\bm{q} = \alpha \bm{u}$ with $\alpha=\frac{16\pi^5 (k_{\mathrm{B}}T)^4}{15(hv_{\mathrm{g}})^3}$~\citep{Guo2015Phonon}. For small temperature differences, the coefficient $\alpha$ can be treated as a constant, giving $\bm{\omega}=\alpha\bm{\nabla}\times\bm{u}$. Another crucial physical quantity,~\textit{the heat dilatation}, is defined as
\begin{equation}
\vartheta\equiv\bm{\nabla}\bm{\cdot}\bm{q}={\rm tr}(\bm{A})={\rm tr}(\bm{D}),
\end{equation}
which characterizes the compressibility of the heat-flux field.

\subsection{Orbital-spin decomposition of heat vorticity}
\begin{figure}[t]
	\centering 
	\includegraphics[width=0.65\columnwidth,trim={8cm 5cm 9cm 5cm},clip]{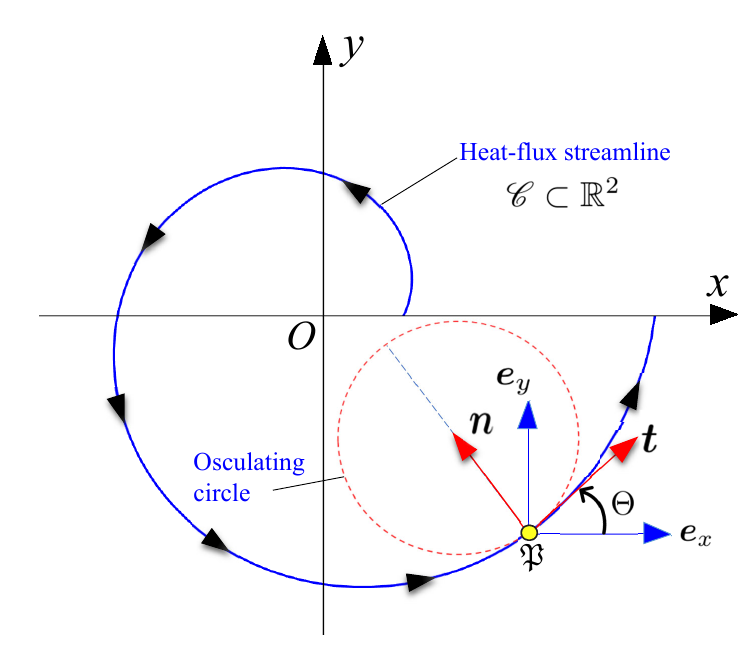}
	\caption{Schematic of the Cartesian frame $(\bm{e}_{1},\bm{e}_{2})$ and the Frenet-Serret frame $(\bm{t},\bm{n})$ along a heat-flux streamline $\mathscr{C}\subset\mathbb{R}^2$. The angle between $\bm{e}_{x}$ and $\bm{t}$ is denoted by $\Theta$. The black arrows indicate the direction of the heat-flux streamline.} 
	\label{Schur_frame_along_pathline}
\end{figure}

Consider a regular $\bm{q}$-streamline $\mathscr{C}\subset\mathbb{R}^2$ oriented by the unit tangent vector $\bm{t}\equiv\bm{q}/\lVert\bm{q}\rVert$, as shown in figure~\ref{Schur_frame_along_pathline}. At each point $\mathfrak{P}$ on $\mathscr{C}$, we introduce a right-handed orthonormal Frenet-Serret triad $(\bm{t},\bm{n},\bm{b})$, where $\bm{n}$ represents the unit principal normal vector (directed toward the local center of curvature), and $\bm{b}\equiv\bm{t}\times\bm{n}$ is the binormal vector perpendicular to the 2D plane~\citep{Frenet1847,Serret1851}. Along $\mathscr{C}$, the heat flux vector is written as $\bm{q}=q\bm{t}$, where $q\equiv\lVert\bm{q}\rVert$ is the heat-flux magnitude. Analogous to previous works for fluid flows~\citep{Chen2025Kinematic,Chen2026General}, a general direction-dependent decomposition for the heat vorticity is proposed as
\begin{subequations}\label{vd}
	\begin{gather}
		\bm{\omega}=\bm{R}(\bm{t})+\bm{S}(\bm{t}), \label{vd1}\\
		\bm{R}(\bm{t})\equiv2\bm{t}\times\left(\bm{t}\bm{\cdot}\bm{A}\right),\quad
		\bm{S}(\bm{t})\equiv-2\bm{t}\times\left(\bm{t}\bm{\cdot}\bm{D}\right), \label{vd2}
	\end{gather}
\end{subequations}
where $\bm{R}(\bm{t})$ and $\bm{S}(\bm{t})$ are referred to as the rigid-rotation mode and the spin mode, respectively. The two heat vorticity modes have clear physical interpretations. The binormal component of $\bm{R}(\bm{t})$ is $\bm{R}(\bm{t})\bm{\cdot}\bm{b}=2\kappa q=\alpha(2\kappa\lVert\bm{u}\rVert)$, which describes a local circular motion with the curvature radius $r_{0}=1/\kappa$ and the speed $\lVert\bm{u}\rVert$, analogous to a speed skater on a curved track. Therefore, $\bm{R}(\bm{t})$ can alternatively be called the orbital-rotation mode. The binormal component of $\bm{S}(\bm{t})$ is $\bm{S}(\bm{t})\bm{\cdot}\bm{b}=(1/r_0)\partial_{r}(rq)=\alpha(1/r_0)\partial_{r}(r\lVert\bm{u}\rVert)$, which arises from the radial gradient of angular momentum, analogous to a figure skater performing a quadruple jump. For a pair of orthogonal material line elements instantaneously aligned with $(\bm{t},\bm{n})$, the rate at which their intersection angle changes (driven by shear deformation) is exactly $\bm{S}(\bm{t})\bm{\cdot}\bm{b}$. This evolution rate is exactly the difference between the angular velocities of the two material line elements.

We define the Cartesian coordinate system $(x,y)$ with the orthonormal basis \((\bm{e}_{x}, \bm{e}_{y}, \bm{e}_{z})\) such that \(\bm{e}_{z} = \bm{e}_{x} \times \bm{e}_{y}\). The components of the vorticity triplet \((\bm{\omega}, \bm{R}, \bm{S})\) along \(\bm{e}_{z}\) are denoted by \((\omega_{z}, R, S)\). Within the \((x,y)\) plane, the heat flux vector is written as \(\bm{q} = q_{x}\bm{e}_{x} + q_{y}\bm{e}_{y}\). The unit tangent vector $\bm{t}=\bm{e}_{x}\cos\Theta+\bm{e}_{y}\sin\Theta$ is calculated as
\begin{equation}
	\cos\Theta = \frac{q_x}{\sqrt{q_x^2 + q_y^2}},\quad
	\sin\Theta = \frac{q_y}{\sqrt{q_x^2 + q_y^2}}.
\end{equation}
The heat-flux gradient tensor $\bm{A}$ and the heat-flux strain-rate tensor $\bm{D}$ are given by
\begin{subequations}\label{VGT_SR_matrices}
	\begin{gather}
		\bm{A}\mapsto\begin{bmatrix}
			A_{xx} & A_{xy} \\
			A_{yx} & A_{yy}
		\end{bmatrix}
		=\begin{bmatrix}
			\partial_{x}q_{x} & \partial_{x}q_{y}\\
			\partial_{y}q_{x} &\partial_{y}q_{y}
		\end{bmatrix},\label{VGT_matrix}\\
		\bm{D}\mapsto\begin{bmatrix}
			D_{xx} & D_{xy} \\
			D_{yx} & D_{yy}
		\end{bmatrix}
		=\begin{bmatrix}
			\partial_{x}q_{x} & \dfrac{1}{2}\left(\partial_{x}q_{y}+\partial_{y}q_{x}\right)\\
			\dfrac{1}{2}\left(\partial_{x}q_{y}+\partial_{y}q_{x}\right) &\partial_{y}q_{y}
		\end{bmatrix}.\label{SR_matrix}
	\end{gather}
\end{subequations}
The vorticity component along $\bm{e}_{z}$ is evaluated as
\begin{equation}\label{vorticity4p6}
	\omega_{z}=A_{xy}-A_{yx}=\partial_{x}q_{y}-\partial_{y}q_{x}.
\end{equation}
From~\eqref{vd1} and~\eqref{vd2}, evaluating their $\bm{e}_{z}$-components gives
\begin{subequations}\label{vorticity4p7abc}
	\begin{gather}
		\omega_{z}=R(\bm{t})+S(\bm{t}), \label{vorticity4p7a}\\
		R(\bm{t})=\langle\bm{t}|\bm{\Phi}_{R}|\bm{t}\rangle=\bm{t}\bm{\cdot}\bm{\Phi}_{R}\bm{\cdot}\bm{t}, \label{vorticity4p7b}\\
		S(\bm{t})=\langle\bm{t}|\bm{\Gamma}_{S}|\bm{t}\rangle=\bm{t}\bm{\cdot}\bm{\Gamma}_{S}\bm{\cdot}\bm{t}. \label{vorticity4p7c}
	\end{gather}
\end{subequations}
Here, $(\bm{\Psi}_{R},\bm{\Gamma}_{S})$ denotes a pair of commutative operators, which can be represented as two real symmetric matrices under any orthonormal basis:
\begin{subequations}\label{vorticity4p8ab}
	\begin{gather}
		\bm{\Phi}_{R}\mapsto\begin{bmatrix}
			2A_{xy} & -(A_{xx}-A_{yy}) \\
			-(A_{xx}-A_{yy}) & -2A_{yx}
		\end{bmatrix},\label{vorticity4p8a}\\
		\bm{\Gamma}_{S}\mapsto\begin{bmatrix}
			-2D_{xy} & D_{xx}-D_{yy} \\
			D_{xx}-D_{yy} & 2D_{yx}
		\end{bmatrix}.\label{vorticity4p8b}
	\end{gather}
\end{subequations}
In the context of vorticity and vortex dynamics, $\bm{\Psi}_{R}$ is termed the rigid-rotation vorticity operator and $\bm{\Gamma}_{S}$ the spin vorticity operator. The flow-field information encoded by these two operators fully characterizes the fundamental morphological
features of vortices, i.e., tubular/axial vortices and winding shear layers with different spin orientations. Their real eigenvalues are precisely the characteristic vorticity modes (which include the rigid-rotation and spin/shear) in the invariant vorticity decompositions, while their complete common eigenvectors just constitute the Schur frame used for the normal-nilpotent decompositions of the velocity gradient tensor~\citep{Chen2026Kinematic,Chen2026Lagrangian}.
\begin{figure}[t]
	\centering
	\includegraphics[width=0.8\columnwidth,trim={4cm 5cm 4cm 4cm},clip]{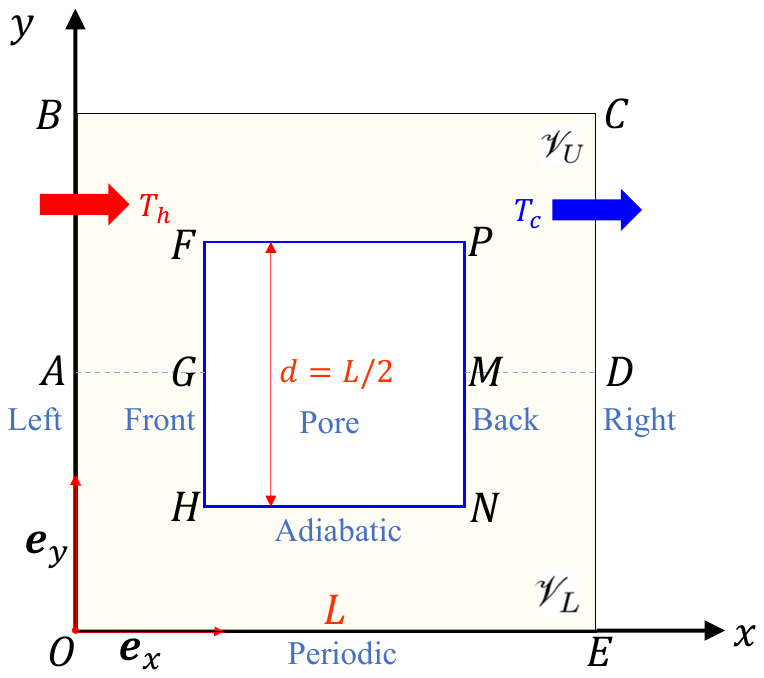}
	\centering
	\caption{Schematic illustration of a unit 2D square cell with a square pore at its center. The side lengths of the unit cell and the square pore are $L$ and $d=L/2$, respectively. The left boundary $\overline{OAB}$ is maintained at a higher temperature $T_{h}$, and the right boundary $\overline{EDC}$ at a lower temperature $T_{c}$. Periodic conditions are applied to the upper and lower boundaries of the unit cell (i.e., $\overline{OE}$ and $\overline{BC}$), whereas adiabatic conditions are used for the boundary $\overline{HFPNH}$ of the square pore. The material domain $\mathscr{V}=\mathscr{V}_{U}\cup\mathscr{V}_{L}$, where $\mathscr{V}_{U}$ and $\mathscr{V}_{L}$ are the upper and lower half regions, respectively.} 
	\label{porous_material}
\end{figure}
\section{Numerical Simulations and Analyses}\label{Numerical simulations and analyses}
\subsection{Problem Description}
Porous media are widely employed in the thermoelectric field, and comprehending their thermal properties is essential for real-world engineering practices. We consider heat conduction in a periodic two-dimensional porous material with aligned square pores. Figure~\ref{porous_material} shows a typical unit square cell with a square pore in the center. The side lengths of the unit cell and the square pore are $L$ and $d=L/2$, respectively. The temperatures of the left and right sides of the cell are $T_{h}=T_{0}+\Delta{T}/2$ and $T_{c}=T_{0}-\Delta{T}/2$, respectively, where the mean temperature $T_{0}=(T_{h}+T_{c})/2$ is chosen as the constant reference temperature, and $\Delta T=T_{h}-T_{c}$ is the temperature difference.
In the ensuing simulations, we set $\Delta{T}/T_{0}=0.01$, which produces a small temperature gradient along the $x$-direction. The temperature in the interior domain is initialized as $T_{0}$. Adiabatic and periodic conditions are applied to the boundaries of the cell and the square pore, respectively. The roughness of the pore surface gives rise to diffuse reflective scattering of phonons when they reach the pore boundaries. The gray model is used. Other parameters are set as $L=2d=1$, $C=1$, $\lVert\bm{v}\rVert=1$, and $T_{0}=1$.

\subsection{Heat Vortices and Vorticity Triplet for Different kinetic regimes}
\begin{figure}[h]
	\centering
	\subfigure[Ballistic, $(\beta_{R},\beta_{N})=(0,0)$]{
		\begin{minipage}[t]{0.33\linewidth}
			\centering
			\includegraphics[width=1.0\columnwidth,trim={0cm 1.3cm 0.0cm 1.3cm},clip]{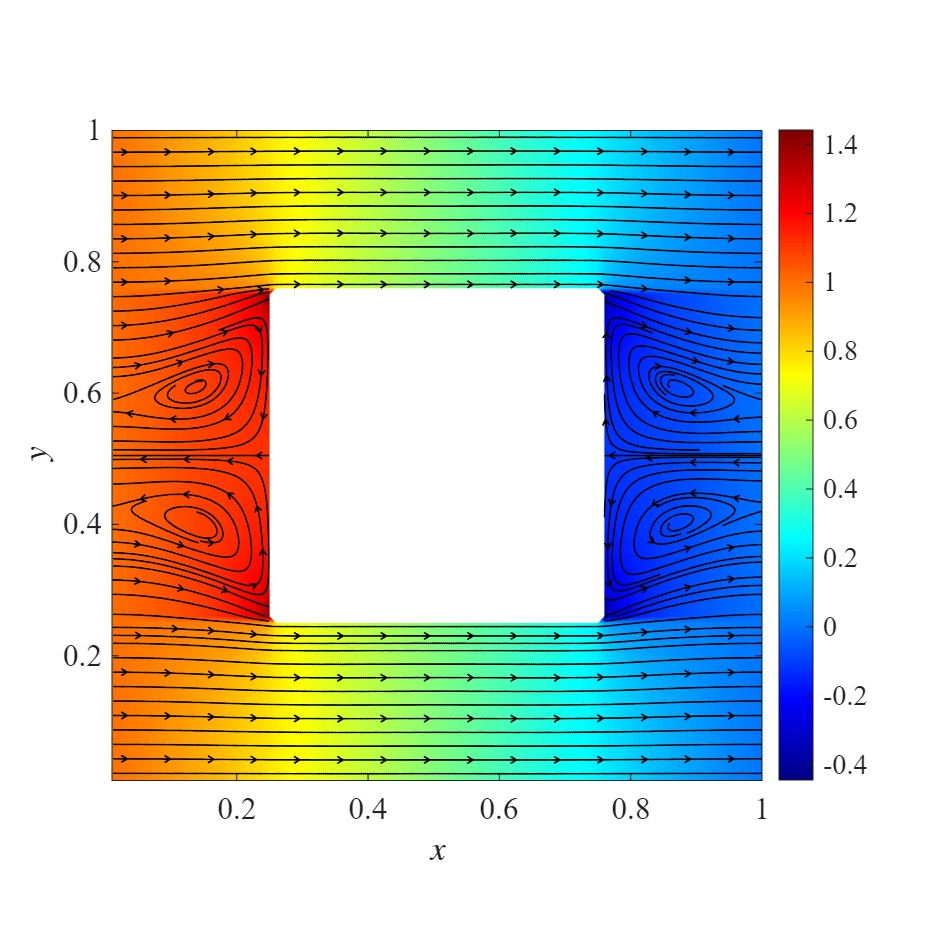}
			\label{T_ballistic}
		\end{minipage}%
	}%
	\subfigure[Diffusive, $(\beta_{R},\beta_{N})=(100,0)$]{
		\begin{minipage}[t]{0.33\linewidth}
			\centering
			\includegraphics[width=1.0\columnwidth,trim={0cm 1.3cm 0.0cm 1.3cm},clip]{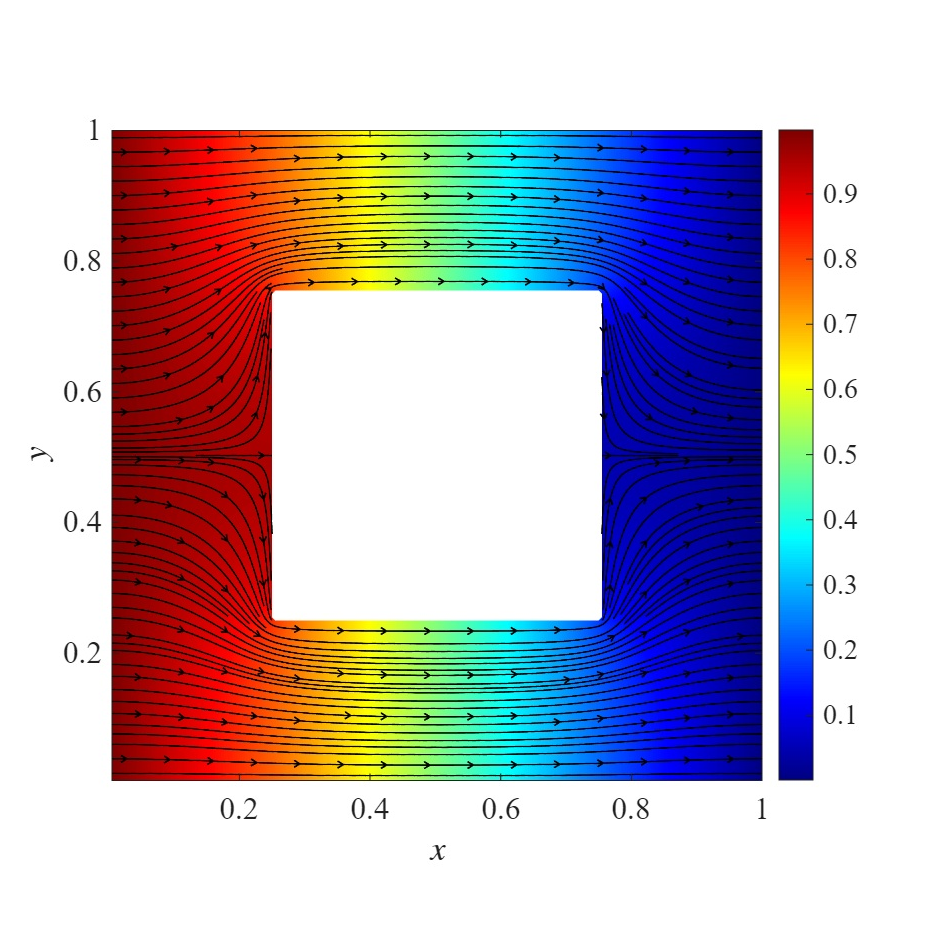}
			\label{T_diffusive}
		\end{minipage}%
	}%
	\subfigure[Hydrodynamic, $(\beta_{R},\beta_{N})=(0,100)$]{
		\begin{minipage}[t]{0.33\linewidth}
			\centering
			\includegraphics[width=1.0\columnwidth,trim={0cm 1.3cm 0.0cm 1.3cm},clip]{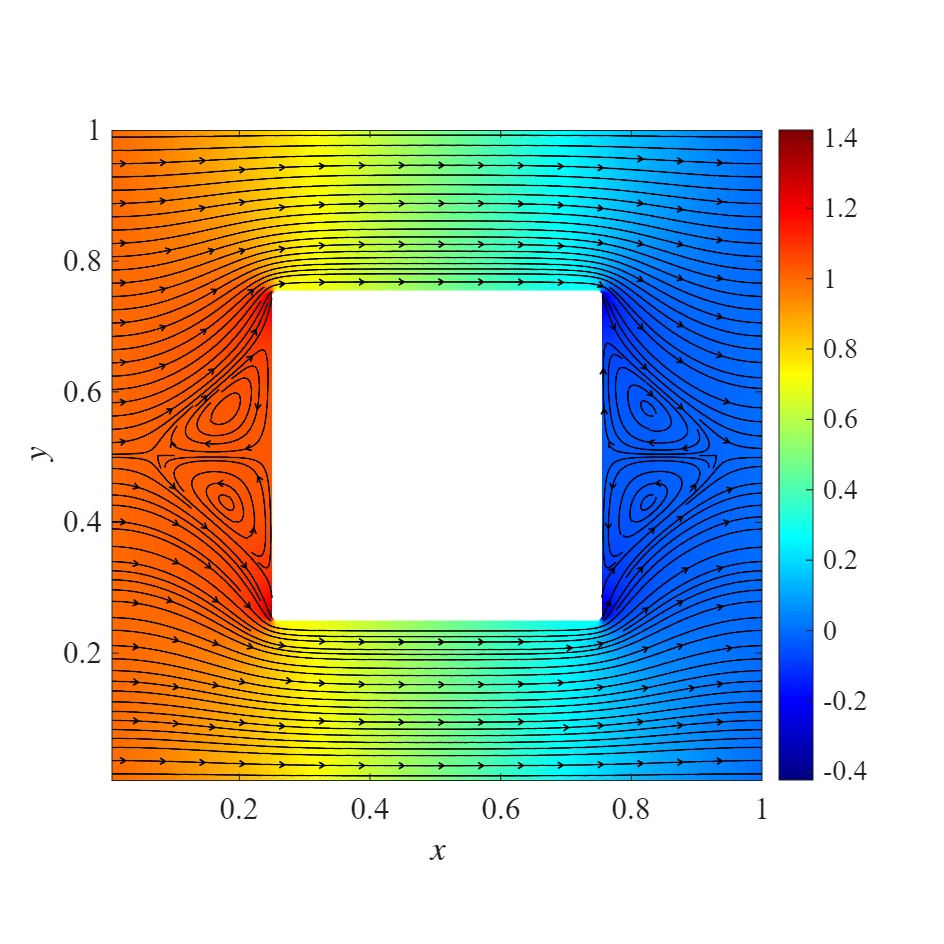}
			\label{T_hydroKn0p01}
		\end{minipage}%
	}%
	
	\subfigure[Hydrodynamic, $(\beta_{R},\beta_{N})=(0,10)$]{
		\begin{minipage}[t]{0.33\linewidth}
			\centering
			\includegraphics[width=1.0\columnwidth,trim={0cm 1.3cm 0.0cm 1.3cm},clip]{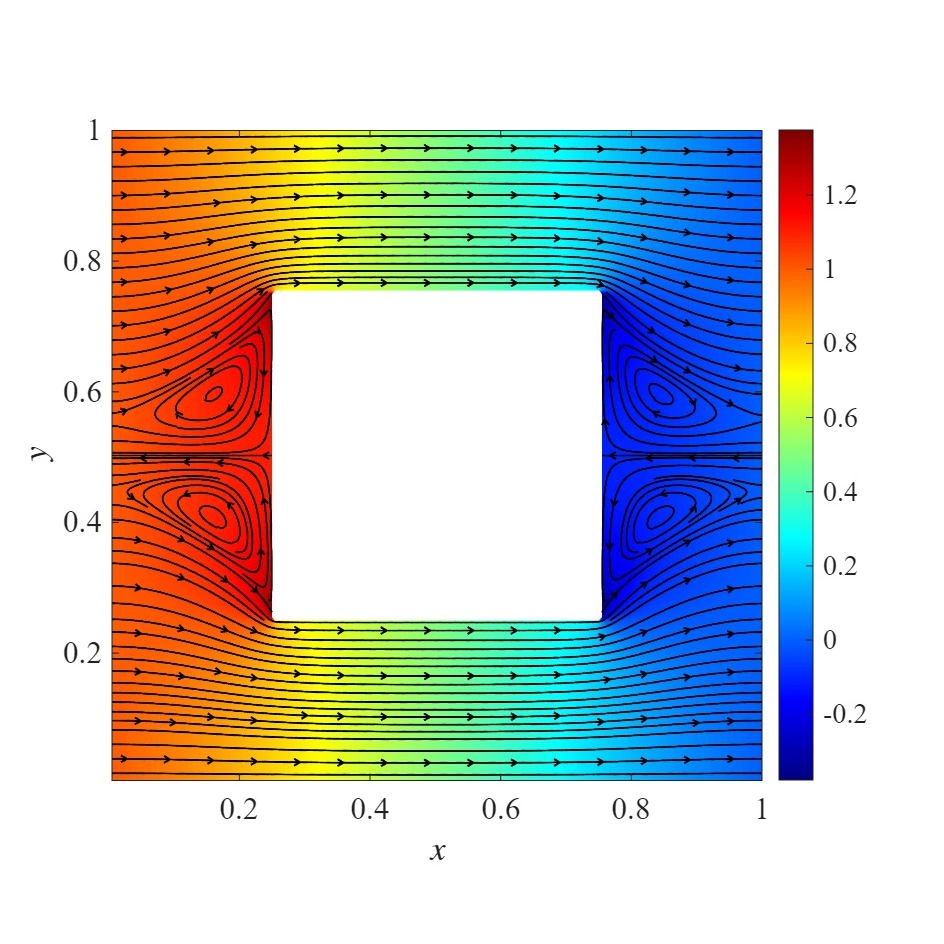}
			\label{T_hydroKn0p1}
		\end{minipage}%
	}%
	\subfigure[Transition, $(\beta_{R},\beta_{N})=(1,100)$]{
		\begin{minipage}[t]{0.33\linewidth}
			\centering
			\includegraphics[width=1.0\columnwidth,trim={0cm 1.3cm 0.0cm 1.3cm},clip]{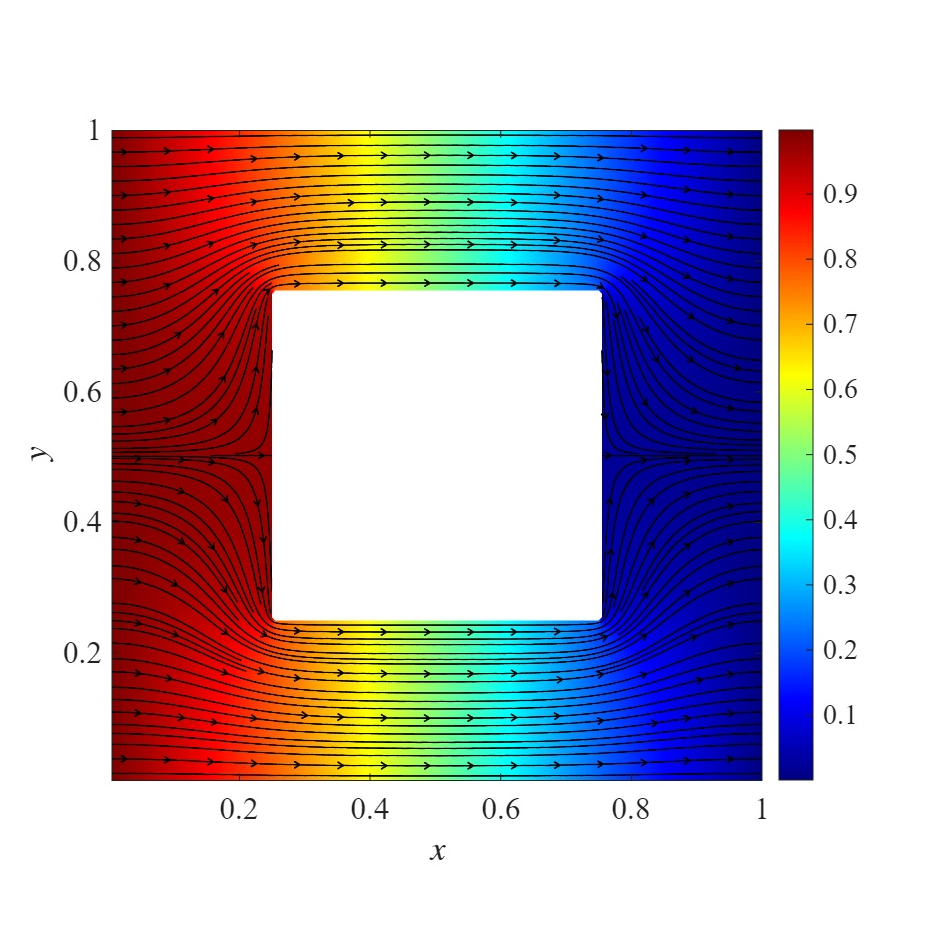}
			\label{T_U1N100}
		\end{minipage}%
	}%
	\subfigure[Transition, $(\beta_{R},\beta_{N})=(1,10)$]{
		\begin{minipage}[t]{0.33\linewidth}
			\centering
			\includegraphics[width=1.0\columnwidth,trim={0cm 1.3cm 0.0cm 1.3cm},clip]{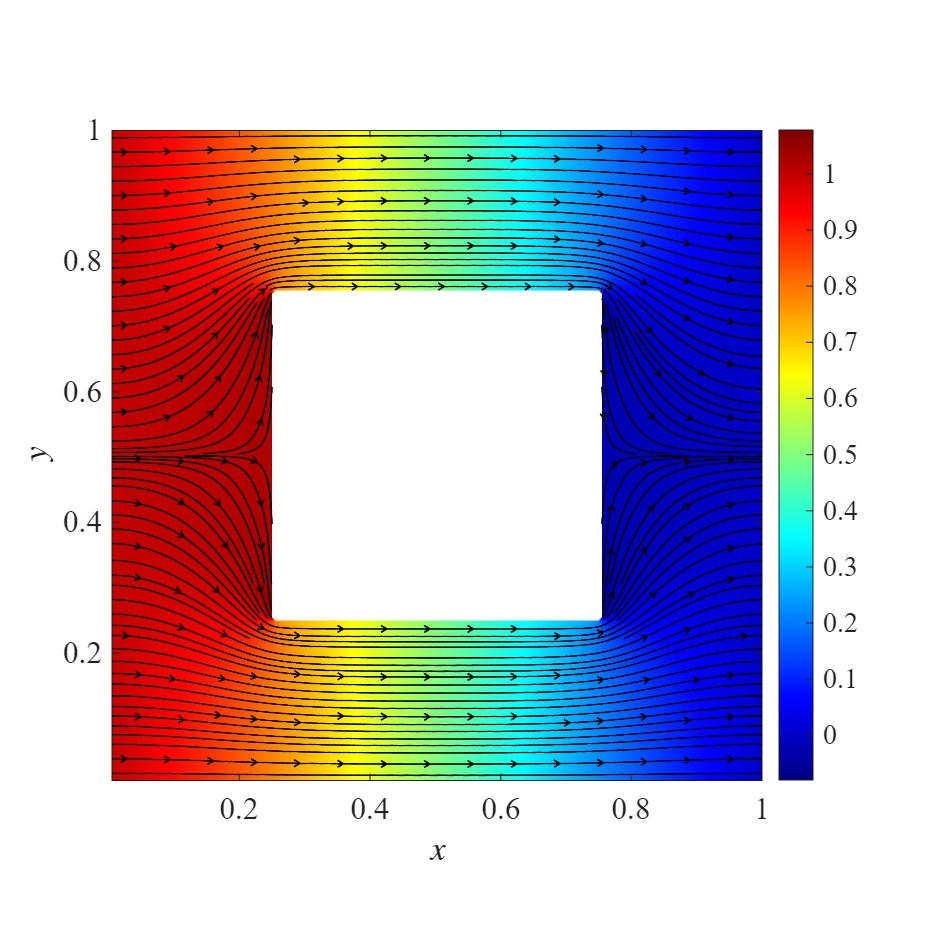}
			\label{T_U1N10}
		\end{minipage}%
	}%
	\caption{Heat flux streamlines superimposed on the normalized contour maps of the temperature $(T-T_c)/\Delta T$ for several limiting regimes. The colorbar are restricted within the range $[-0.1,0.1]$.}
	\label{Temperature}
\end{figure}
\begin{figure}[h!]
	\centering
	\subfigure[Ballistic, $(\beta_{R},\beta_{N})=(0,0)$]{
		\begin{minipage}[t]{0.33\linewidth}
			\centering
			\includegraphics[width=1.0\columnwidth,trim={0cm 1.3cm 0.0cm 1.3cm},clip]{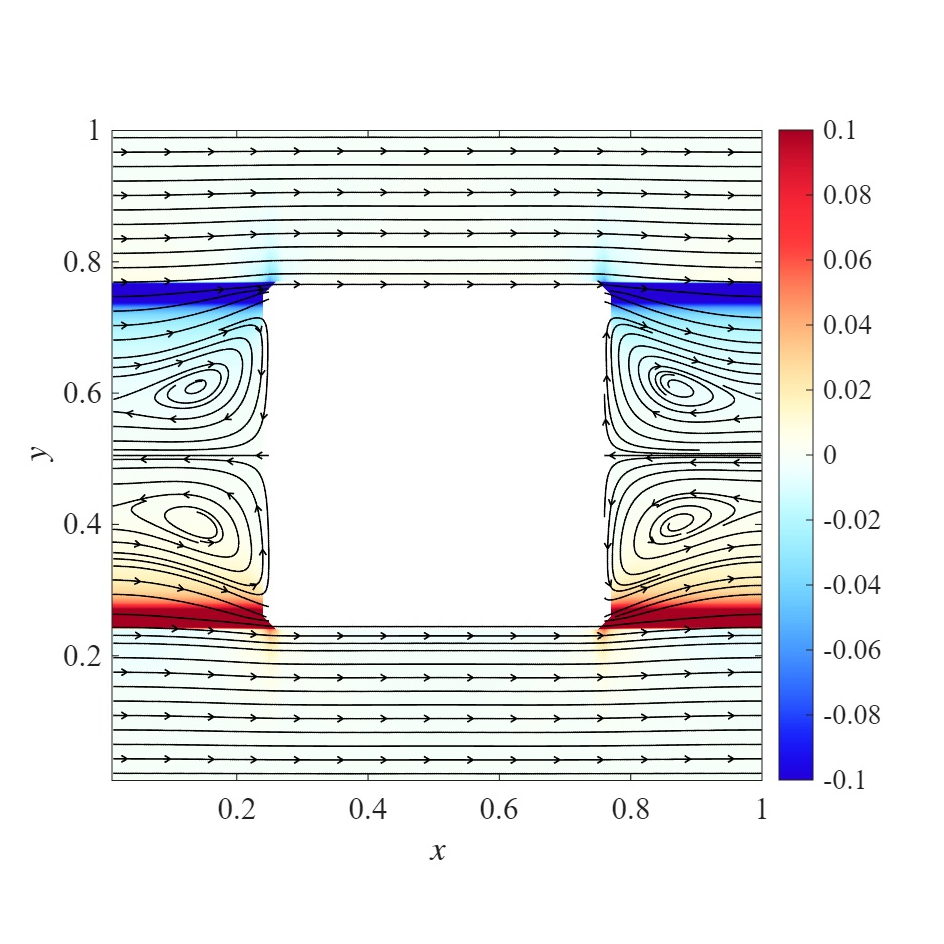}
			\label{Vorticity_ballistic}
		\end{minipage}%
	}%
	\subfigure[Diffusive, $(\beta_{R},\beta_{N})=(100,0)$]{
		\begin{minipage}[t]{0.33\linewidth}
			\centering
			\includegraphics[width=1.0\columnwidth,trim={0cm 1.3cm 0.0cm 1.3cm},clip]{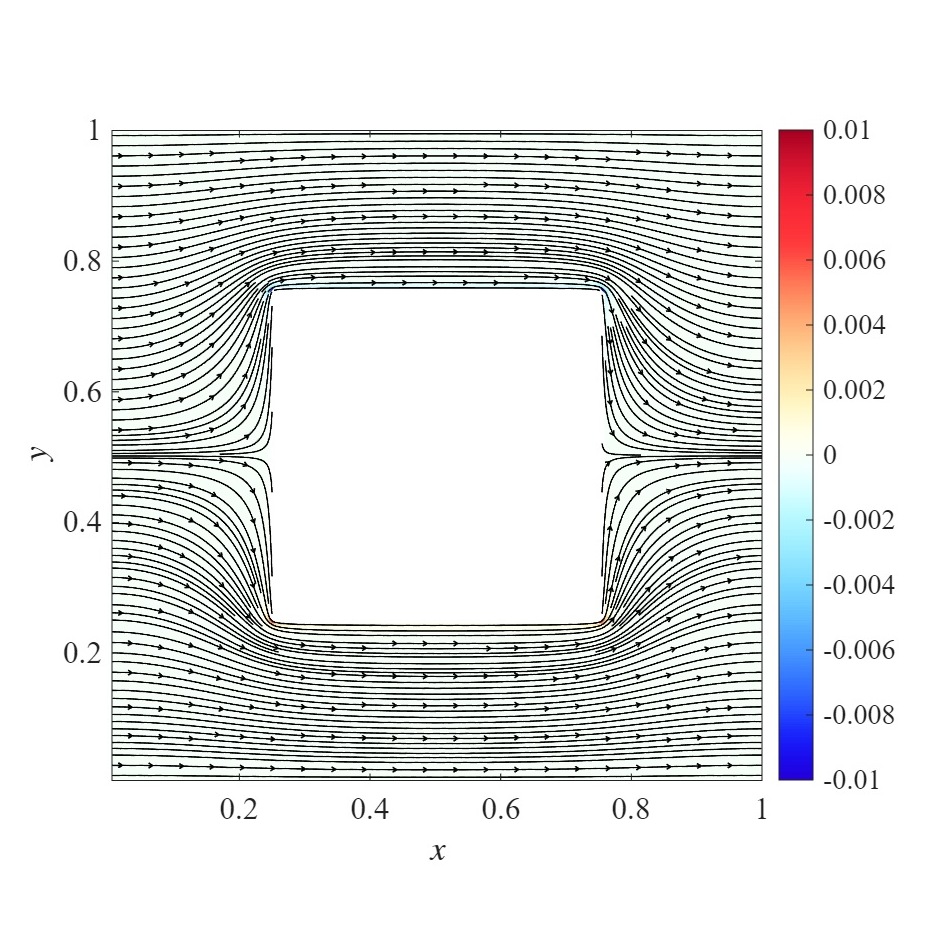}
			\label{Vorticity_diffusive}
		\end{minipage}%
	}%
	\subfigure[Hydrodynamic, $(\beta_{R},\beta_{N})=(0,100)$]{
		\begin{minipage}[t]{0.33\linewidth}
			\centering
			\includegraphics[width=1.0\columnwidth,trim={0cm 1.3cm 0.0cm 1.3cm},clip]{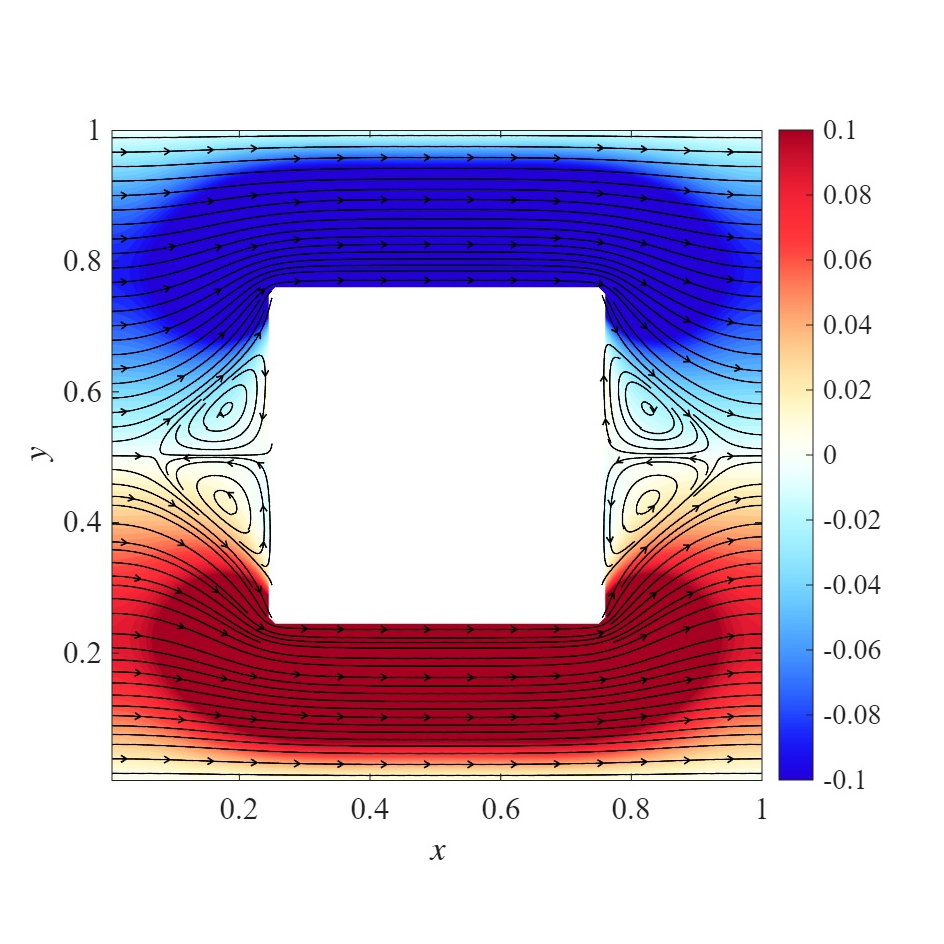}
			\label{Vorticity_hydroKn0p01}
		\end{minipage}%
	}%
	
	\subfigure[Hydrodynamic, $(\beta_{R},\beta_{N})=(0,10)$]{
		\begin{minipage}[t]{0.33\linewidth}
			\centering
			\includegraphics[width=1.0\columnwidth,trim={0cm 1.3cm 0.0cm 1.3cm},clip]{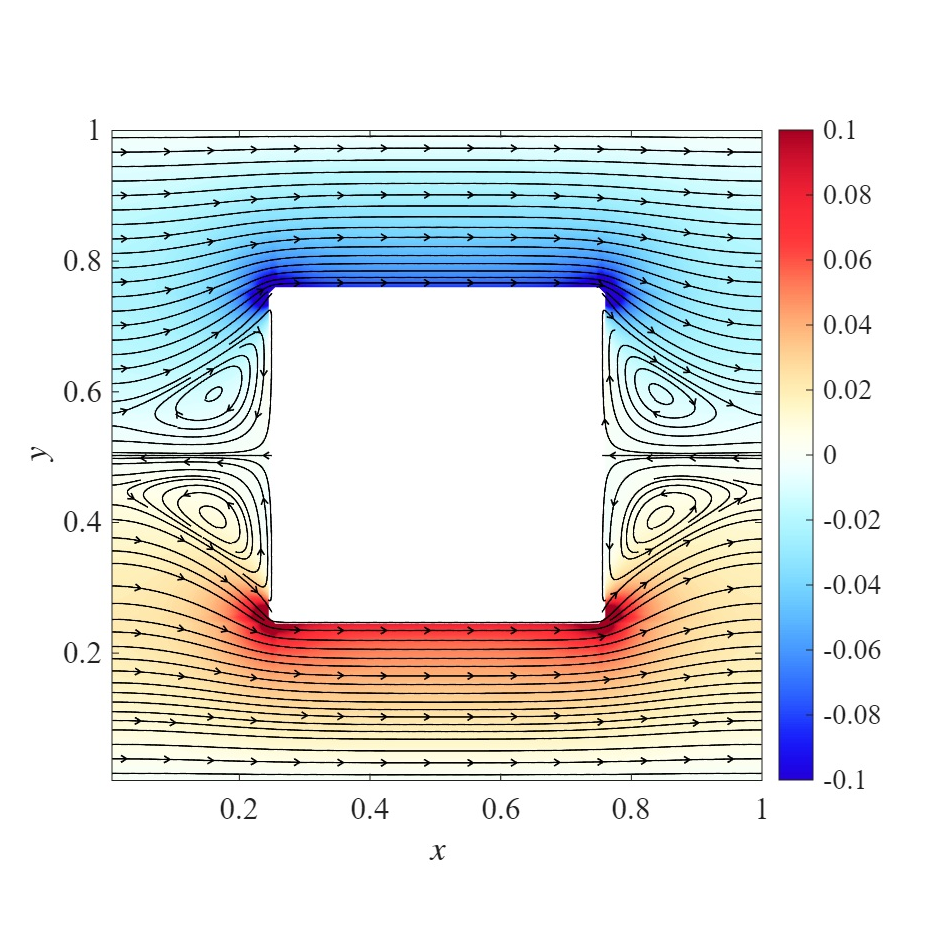}
			\label{Vorticity_hydroKn0p1}
		\end{minipage}%
	}%
	\subfigure[Transition, $(\beta_{R},\beta_{N})=(1,100)$]{
		\begin{minipage}[t]{0.33\linewidth}
			\centering
			\includegraphics[width=1.0\columnwidth,trim={0cm 1.3cm 0.0cm 1.3cm},clip]{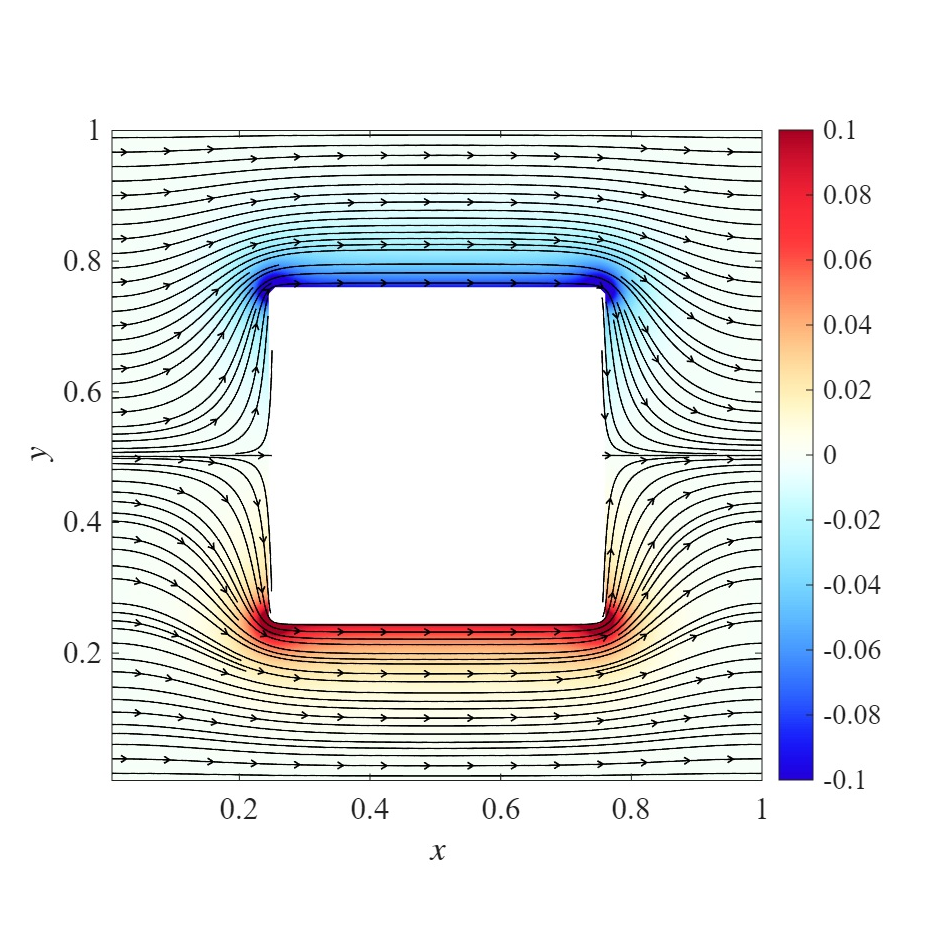}
			\label{omegaz_U1N100}
		\end{minipage}%
	}%
	\subfigure[Transition, $(\beta_{R},\beta_{N})=(1,10)$]{
		\begin{minipage}[t]{0.33\linewidth}
			\centering
			\includegraphics[width=1.0\columnwidth,trim={0cm 1.3cm 0.0cm 1.3cm},clip]{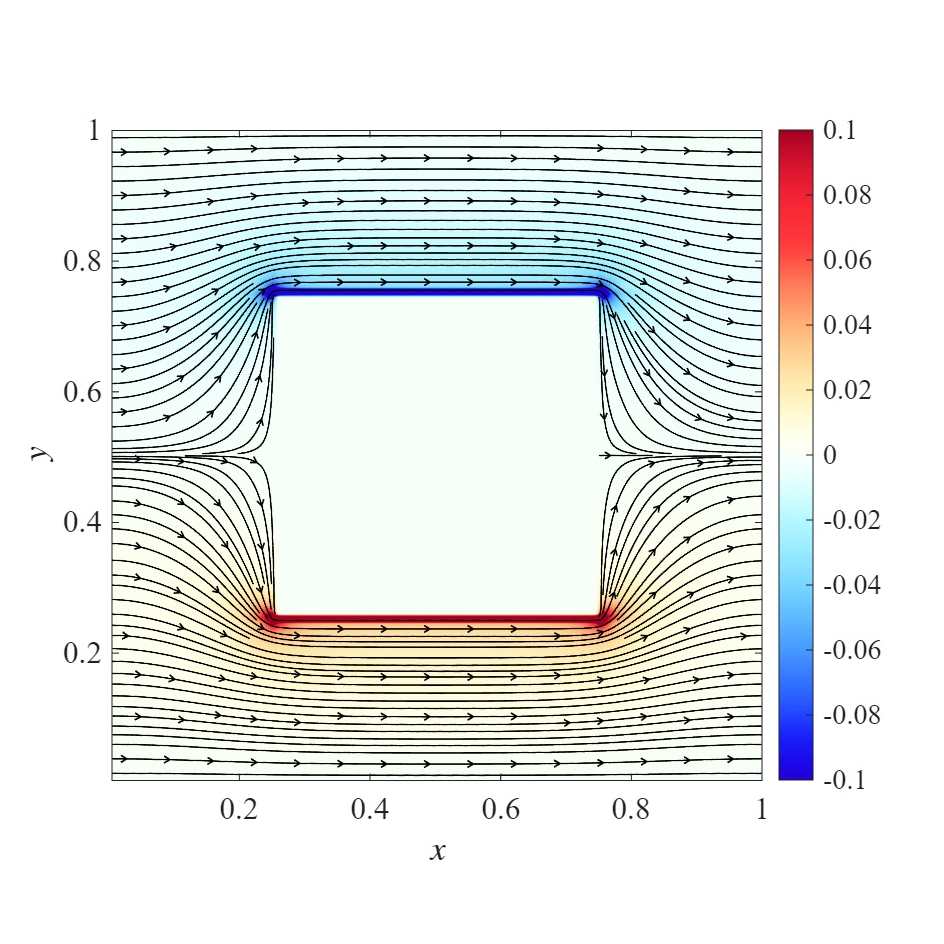}
			\label{omegaz_U1N10}
		\end{minipage}%
	}%
	\caption{Heat flux streamlines superimposed on the normalized contour maps of the vorticity $\omega_{z}$ for several limiting regimes. The colorbar are restricted within the range $[-0.1,0.1]$.}
	\label{Vorticity_limiting_regimes}
\end{figure}
\begin{figure}[t]
	\centering
	\subfigure[Ballistic, $(\beta_{R},\beta_{N})=(0,0)$]{
		\begin{minipage}[t]{0.33\linewidth}
			\centering
			\includegraphics[width=1.0\columnwidth,trim={0cm 1.3cm 0.0cm 1.3cm},clip]{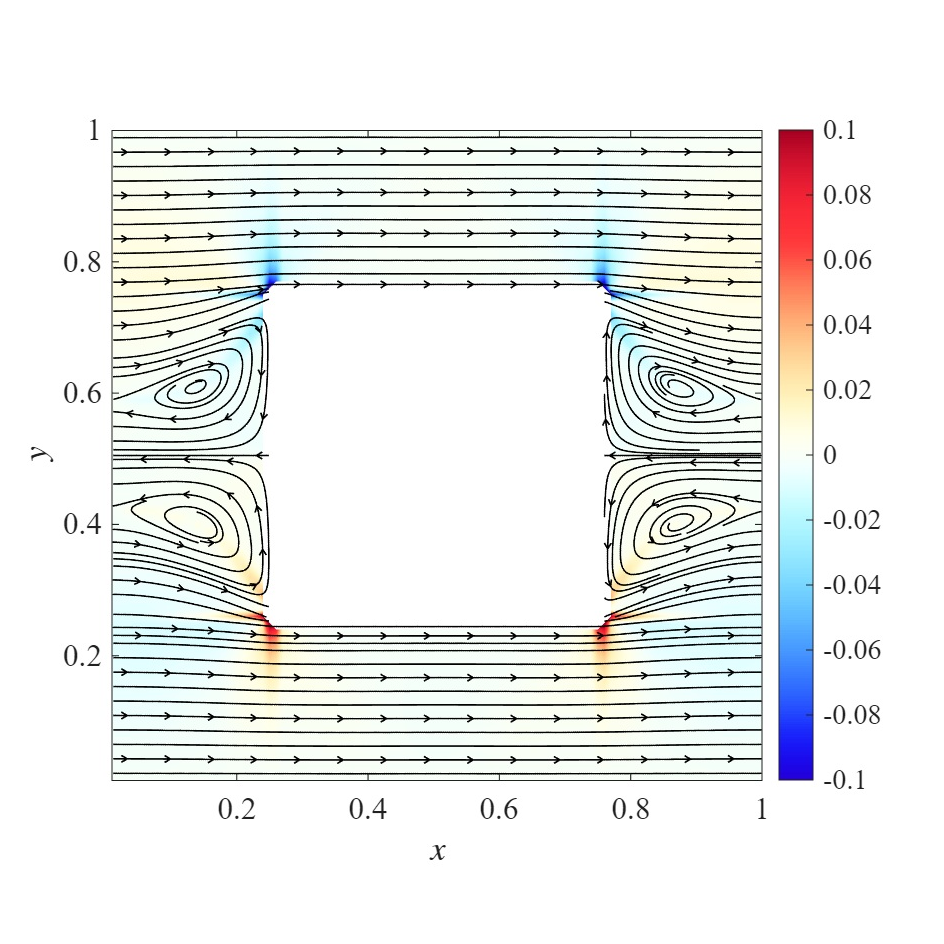}
			\label{Ballistic}
		\end{minipage}%
	}%
	\subfigure[Diffusive, $(\beta_{R},\beta_{N})=(100,0)$]{
		\begin{minipage}[t]{0.33\linewidth}
			\centering
			\includegraphics[width=1.0\columnwidth,trim={0cm 1.3cm 0.0cm 1.3cm},clip]{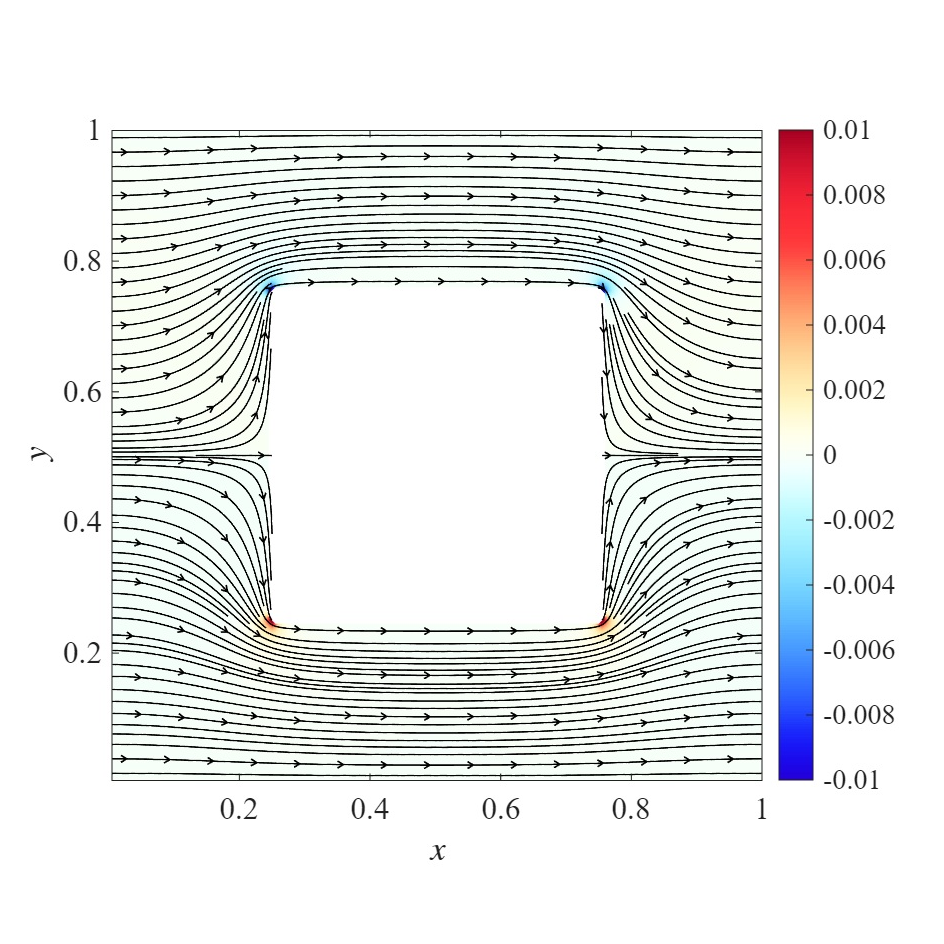}
			\label{diffusive}
		\end{minipage}%
	}%
	\subfigure[Hydrodynamic, $(\beta_{R},\beta_{N})=(0,100)$]{
		\begin{minipage}[t]{0.33\linewidth}
			\centering
			\includegraphics[width=1.0\columnwidth,trim={0cm 1.3cm 0.0cm 1.3cm},clip]{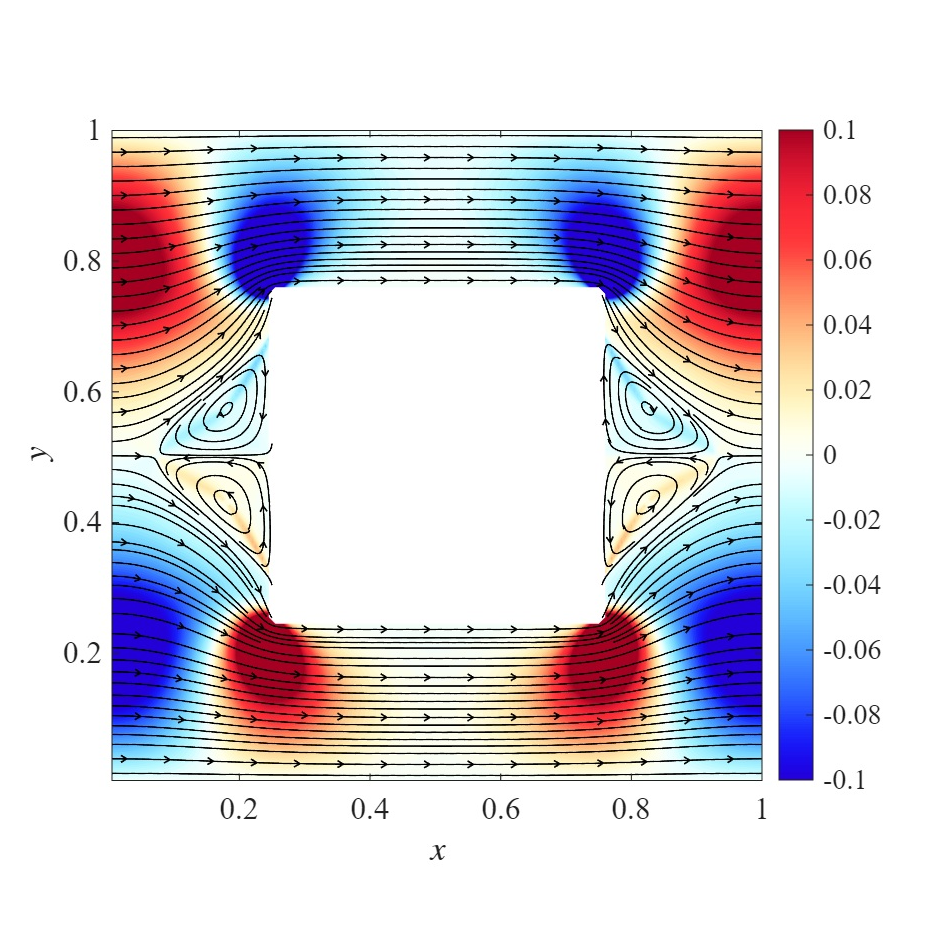}
			\label{Rmode_hydroKn0p01}
		\end{minipage}%
	}%
	
	\subfigure[Hydrodynamic, $(\beta_{R},\beta_{N})=(0,10)$]{
		\begin{minipage}[t]{0.33\linewidth}
			\centering
			\includegraphics[width=1.0\columnwidth,trim={0cm 1.3cm 0.0cm 1.3cm},clip]{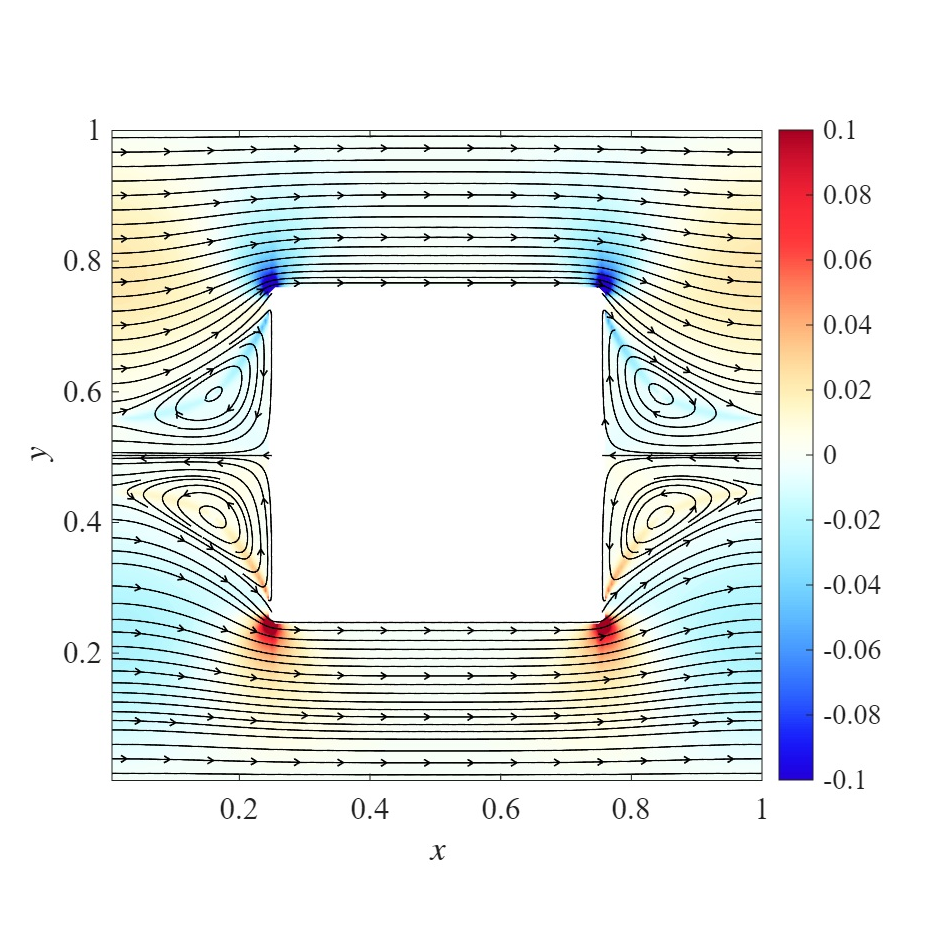}
			\label{Rmode_hydroKn0p1}
		\end{minipage}%
	}%
	\subfigure[Transition, $(\beta_{R},\beta_{N})=(1,100)$]{
		\begin{minipage}[t]{0.33\linewidth}
			\centering
			\includegraphics[width=1.0\columnwidth,trim={0cm 1.3cm 0.0cm 1.3cm},clip]{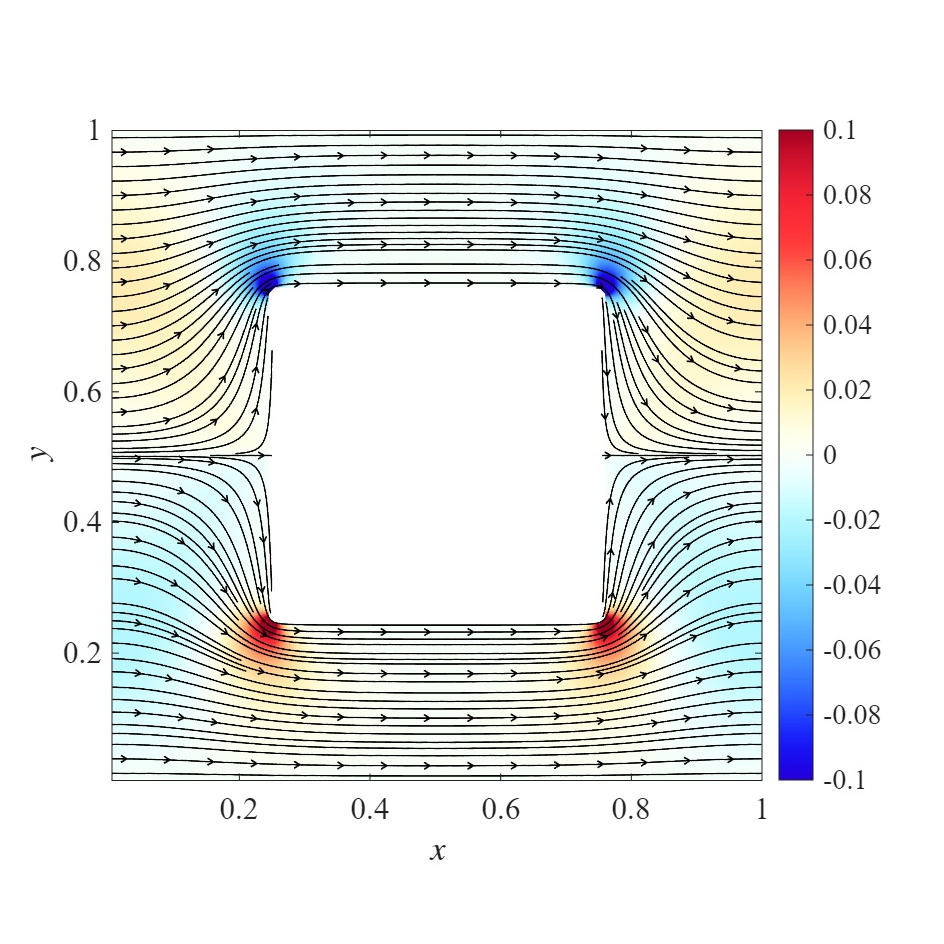}
			\label{Rmode_U1N100}
		\end{minipage}%
	}%
	\subfigure[Transition, $(\beta_{R},\beta_{N})=(1,10)$]{
		\begin{minipage}[t]{0.33\linewidth}
			\centering
			\includegraphics[width=1.0\columnwidth,trim={0cm 1.3cm 0.0cm 1.3cm},clip]{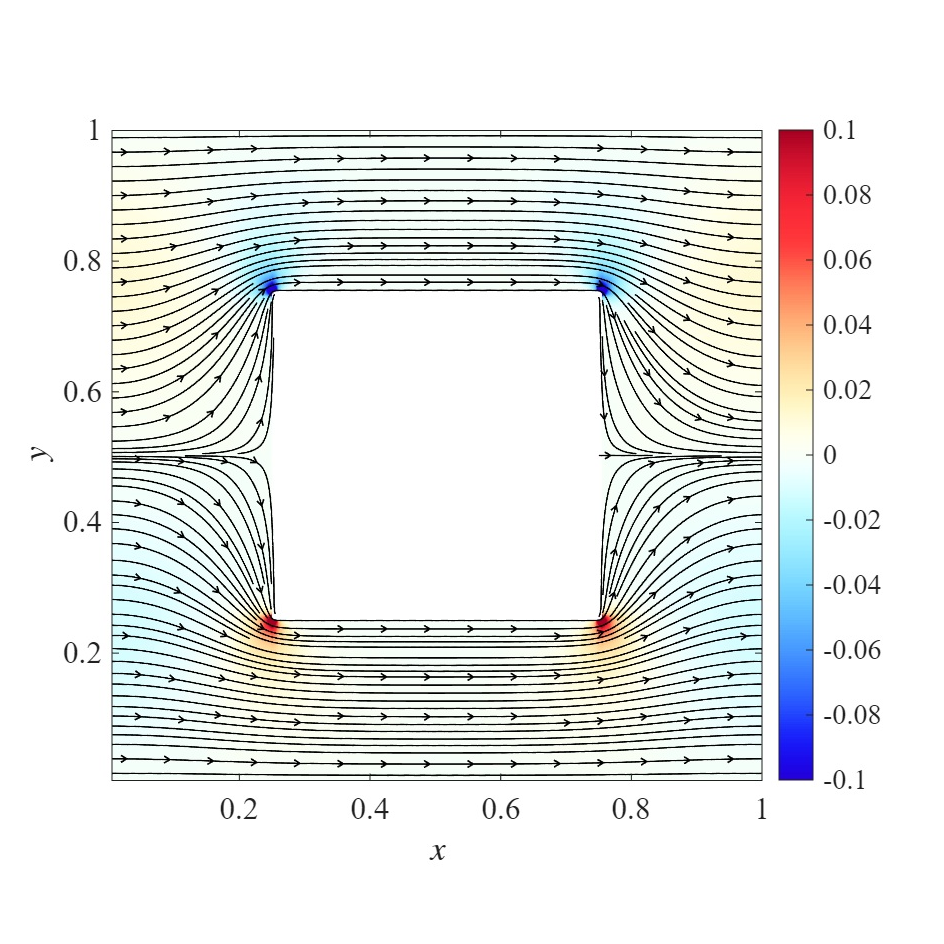}
			\label{Rmode_U1N10}
		\end{minipage}%
	}%
	\caption{Heat flux streamlines superimposed on the normalized contour maps of the orbital-rotation mode ${R}(\bm{t})$ for several limiting regimes. The colorbar are restricted within the range $[-0.1,0.1]$.}
	\label{Rmode_limiting_regimes}
\end{figure}
\begin{figure}[h!]
	\centering
	\subfigure[Ballistic, $(\beta_{R},\beta_{N})=(0,0)$]{
		\begin{minipage}[t]{0.33\linewidth}
			\centering
			\includegraphics[width=1.0\columnwidth,trim={0cm 1.3cm 0.0cm 1.3cm},clip]{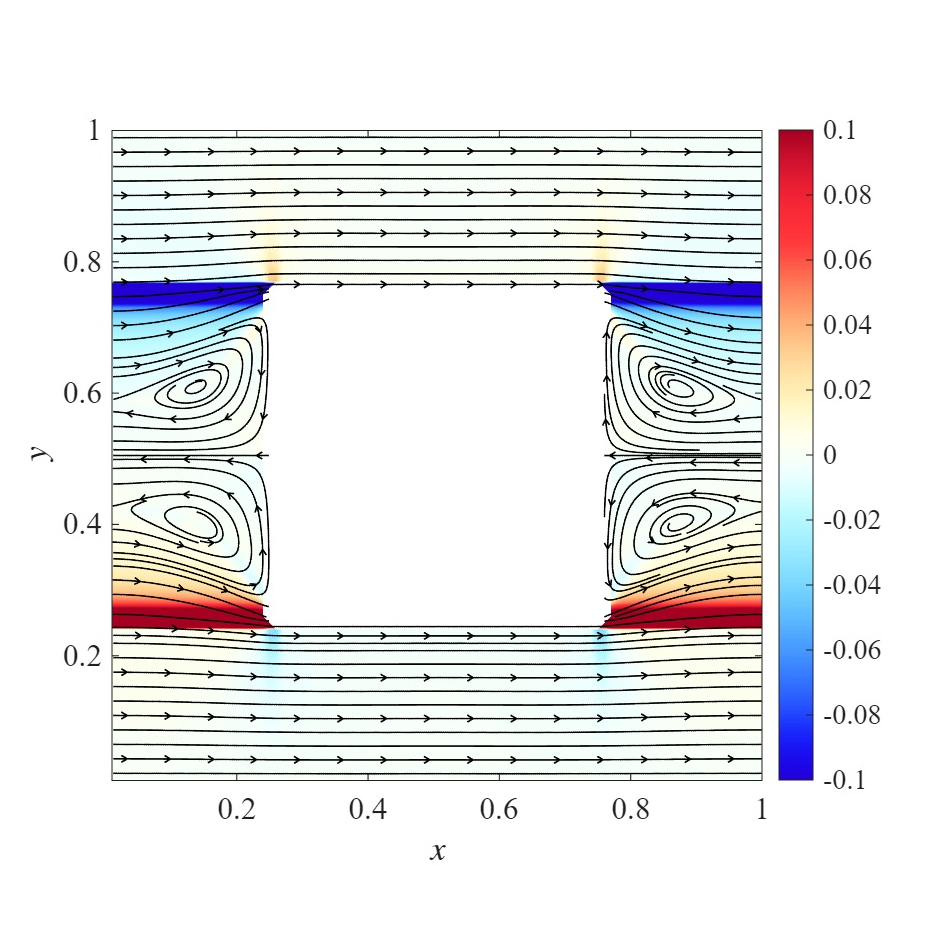}
			\label{Smode_ballistic}
		\end{minipage}%
	}%
	\subfigure[Diffusive, $(\beta_{R},\beta_{N})=(100,0)$]{
		\begin{minipage}[t]{0.33\linewidth}
			\centering
			\includegraphics[width=1.0\columnwidth,trim={0cm 1.3cm 0.0cm 1.3cm},clip]{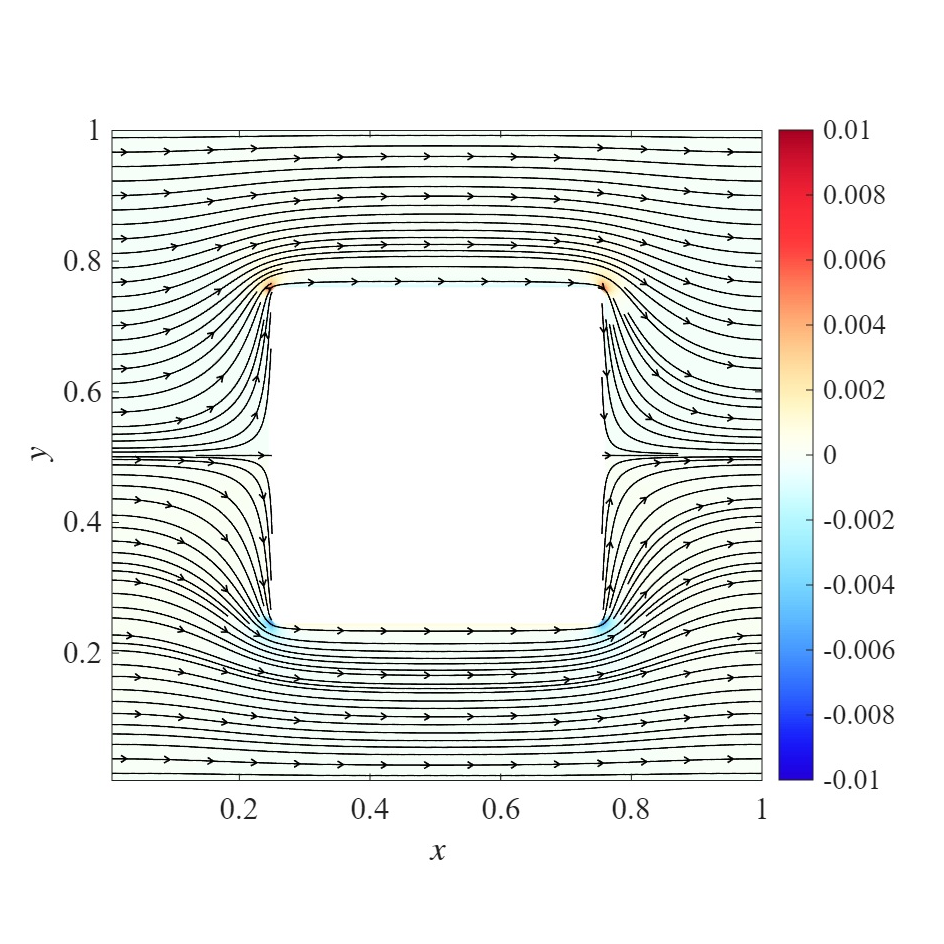}
			\label{Smode_diffusive}
		\end{minipage}%
	}%
	\subfigure[Hydrodynamic, $(\beta_{R},\beta_{N})=(0,100)$]{
		\begin{minipage}[t]{0.33\linewidth}
			\centering
			\includegraphics[width=1.0\columnwidth,trim={0cm 1.3cm 0.0cm 1.3cm},clip]{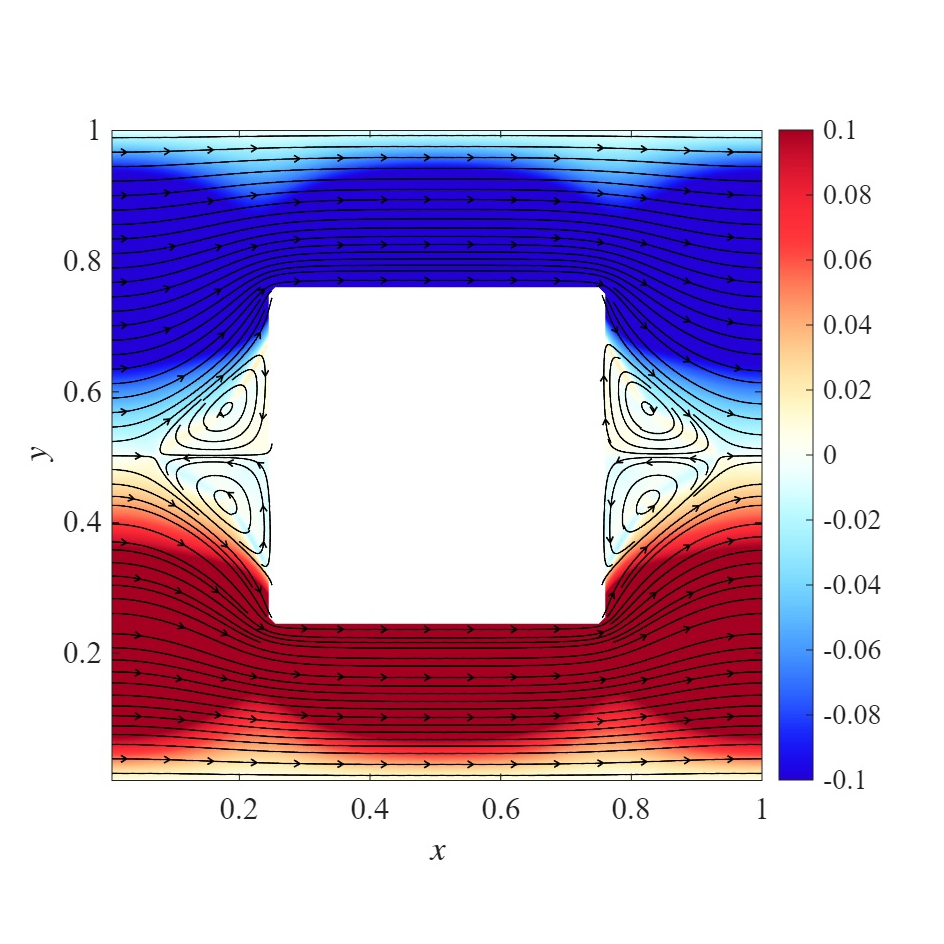}
			\label{Smode_hydroKn0p01}
		\end{minipage}%
	}%
	
	\subfigure[Hydrodynamic, $(\beta_{R},\beta_{N})=(0,10)$]{
		\begin{minipage}[t]{0.33\linewidth}
			\centering
			\includegraphics[width=1.0\columnwidth,trim={0cm 1.3cm 0.0cm 1.3cm},clip]{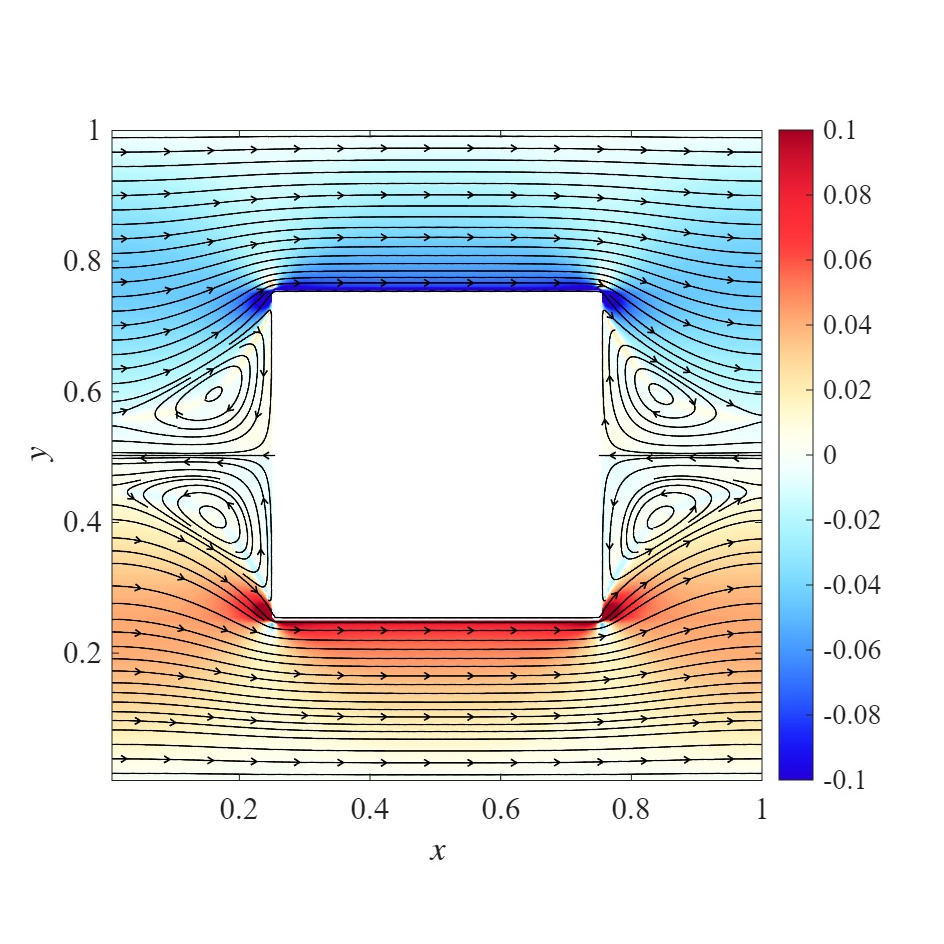}
			\label{Smode_hydroKn0p1}
		\end{minipage}%
	}%
	\subfigure[Transition, $(\beta_{R},\beta_{N})=(1,100)$]{
		\begin{minipage}[t]{0.33\linewidth}
			\centering
			\includegraphics[width=1.0\columnwidth,trim={0cm 1.3cm 0.0cm 1.3cm},clip]{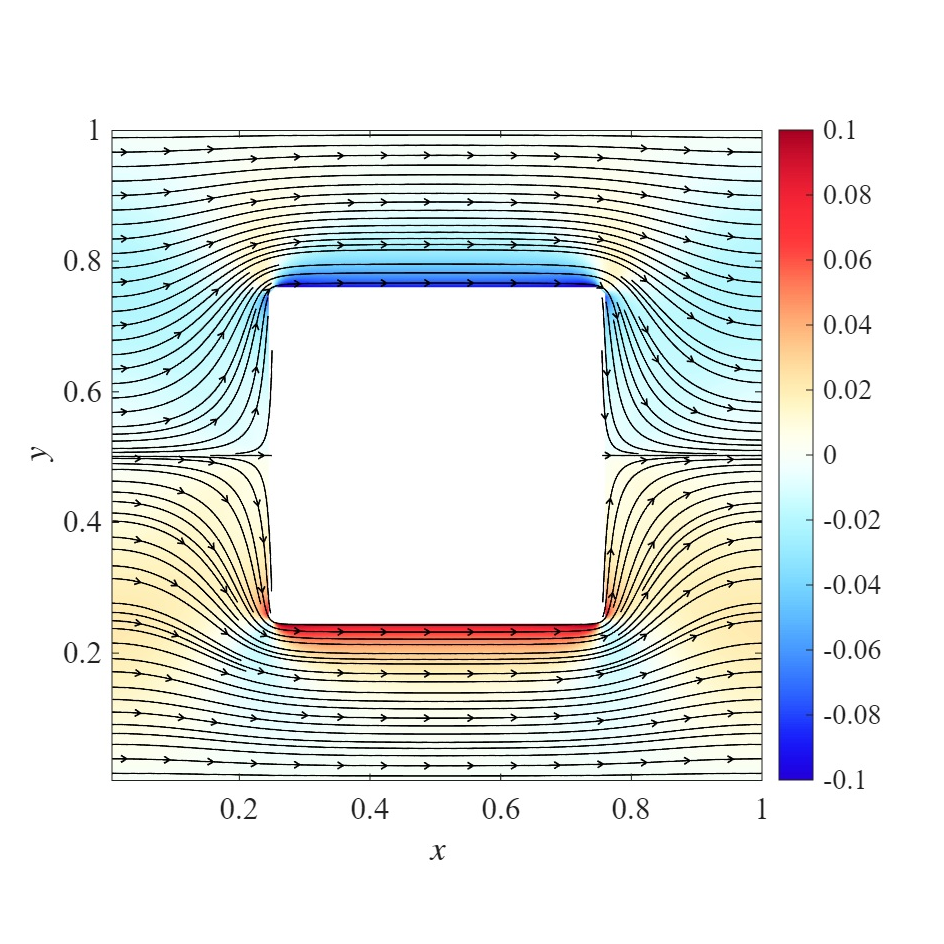}
			\label{Smode_U1N100}
		\end{minipage}%
	}%
	\subfigure[Transition, $(\beta_{R},\beta_{N})=(1,10)$]{
		\begin{minipage}[t]{0.33\linewidth}
			\centering
			\includegraphics[width=1.0\columnwidth,trim={0cm 1.3cm 0.0cm 1.3cm},clip]{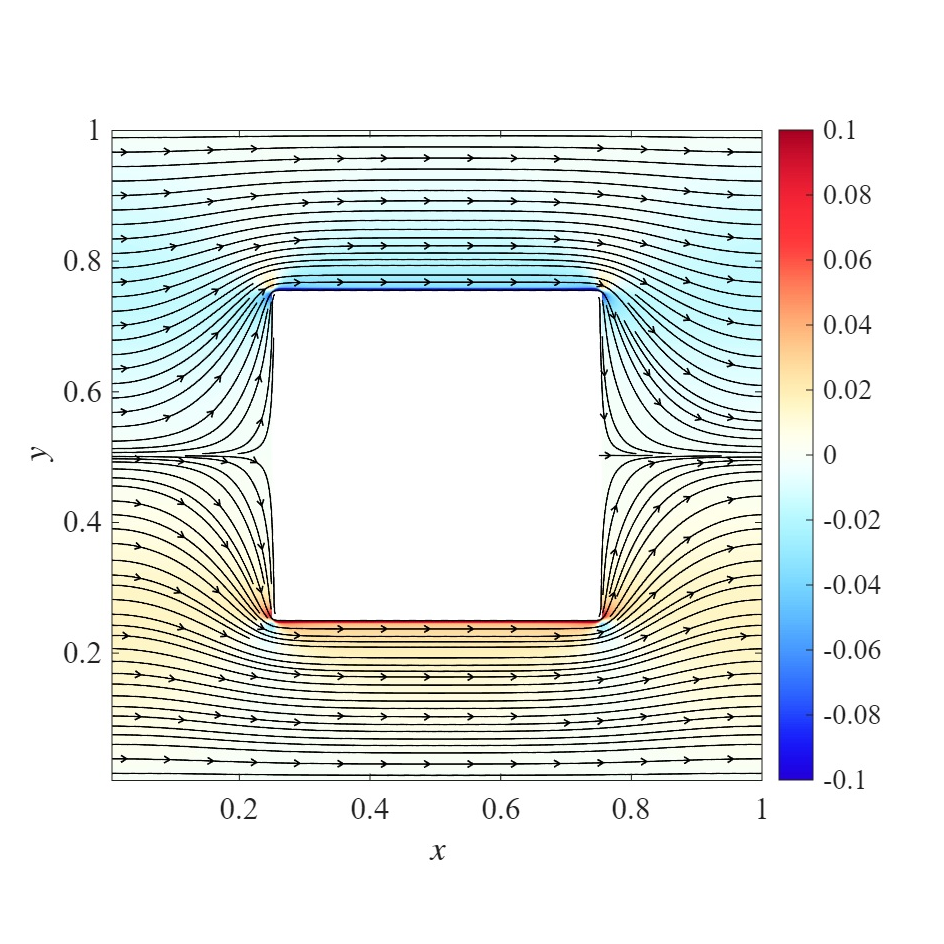}
			\label{Smode_U1N10}
		\end{minipage}%
	}%
	\caption{Heat flux streamlines superimposed on the normalized contour maps of the spin mode $S(\bm{t})$ for several limiting regimes. The colorbar are restricted within the range $[-0.1,0.1]$.}
	\label{Smode_limiting_regimes}
\end{figure}

Figure~\ref{Temperature} shows the heat-flux streamlines superimposed on the contour maps of the normalized temperature for the ballistic, diffusive, hydrodynamic, and transition regimes. Correspondingly, figure~\ref{Vorticity_limiting_regimes} displays the contour plots of the heat vorticity field, while those of the orbital-rotation and spin modes are presented in figures~\ref{Rmode_limiting_regimes} and~\ref{Smode_limiting_regimes}, respectively. Figures~\ref{compare_abc} and~\ref{compare_abc1} provide a quantitative comparison of the vorticity triplet $(\bm{R},\bm{S},\bm{\omega})$ along the straight line connecting the centers of the heat vortices for different kinetic regimes.

The ballistic limit is defined by $(\beta_{R},\beta_{N})=(0,0)$, or equivalently, $(\tau_{R},\tau_{N})=(\infty,\infty)$, which corresponds to the regime of rare phonon scattering. In this regime, thermal conduction is governed by phonon convection. According to~\eqref{BTE2} (i.e., $\partial_{t}e+\bm{v}\bm{\cdot}\bm{\nabla}e=0$), the phonon deviational energy distribution function remains invariant when traveling at the group velocity $\bm{v}$; however, it can be modified by boundary scattering. Since phonon relaxation effects are negligible,~\eqref{eq29b} (i.e., $\partial_{t}\bm{q}+\bm{\nabla}\bm{\cdot}\bm{Q}=\bm{0}$) implies that the total heat flux within the entire domain must be conserved, while the distribution of the heat-flux field is affected by the specified boundary conditions. 

As shown in figure~\ref{T_ballistic}, for $0.75<y<1$, heat propagates rightward, driven by the weak temperature gradient along the $x$-direction. Upstream of the square pole $(0.25<y<0.75,~0<x<0.25)$, right-traveling phonons strike the pore boundary and undergo diffuse reflection. The accumulation of high-energy phonons impinging on and bouncing off the front pore boundary produces an elevated temperature immediately upstream of the pore. Similar physical reasoning accounts for the reduced temperature observed downstream of the pore. Under the diffuse boundary condition, phonons emitted from the pore boundary experience diffuse reflection, which induces a leftward heat flux in the neighborhood of the symmetry line $(y=0.5)$. Consequently, in figure~\ref{Vorticity_ballistic}, the interaction between this reverse heat flux and the incoming stream leads to the formation of heat vortices that induce high vorticity magnitudes nearby the parallel lines $y=0.25$ and $0.75$. In figures~\ref{Ballistic} and~\ref{Smode_ballistic}, it is seen that the orbital rotation $R$ is mainly concentrated near the pole corners, and the vorticity pattern is dominated by the spin mode $S$. These observations are further supported by the quantitative comparison in figure~\ref{compare_Ballistic}. Along the straight line connecting the centers of the heat vortices, the contribution of $R$ to $\omega_{z}$ is negligible whereas $S$ exhibits two sharp peaks around $y=0.25$ and $0.75$ due to the interaction between the outer heat flow and the heat vortices.

\begin{figure}[htbp]
	\centering
	\subfigure[Ballistic regime, $(\beta_{R},\beta_{N})=(0,0)$]{
		\begin{minipage}[t]{1.0\linewidth}
			\centering
			\includegraphics[width=1.0\columnwidth,trim={0cm 5cm 0cm 4cm},clip]{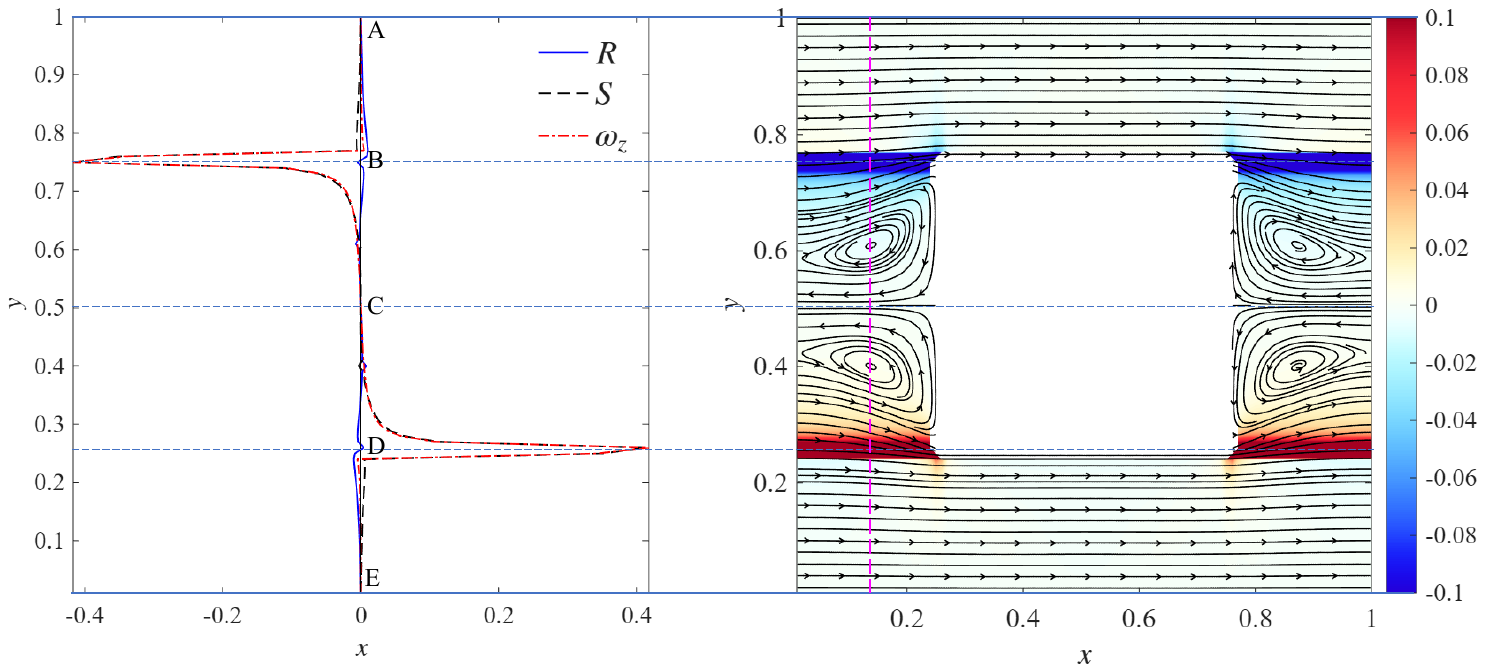}
			\label{compare_Ballistic}
		\end{minipage}%
	}%
	
	\subfigure[Diffusive regime, $(\beta_{R},\beta_{N})=(0.05,100)$]{
		\begin{minipage}[t]{1.0\linewidth}
			\centering
			\includegraphics[width=1.0\columnwidth,trim={0cm 5cm 0cm 4cm},clip]{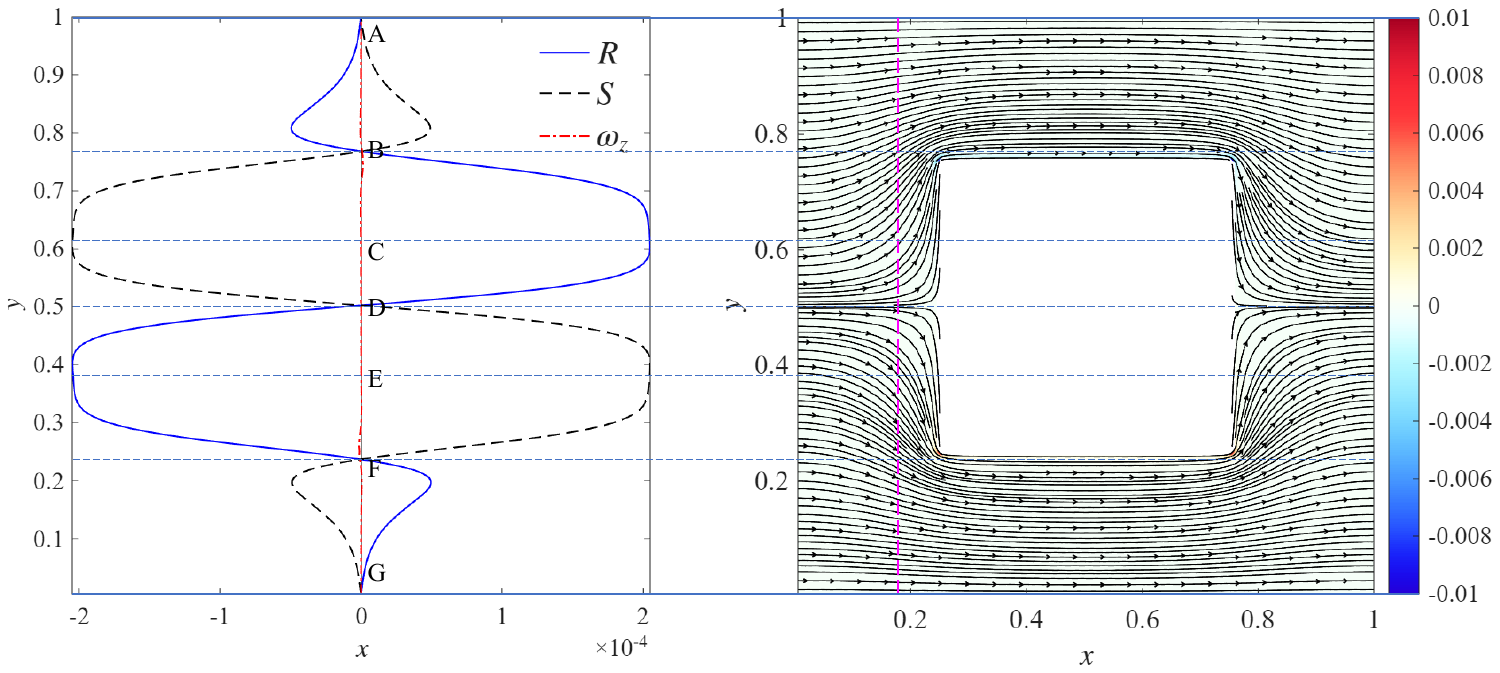}
			\label{compare_Diffuse}
		\end{minipage}%
	}%
	\caption{Comparison of the spatial distributions of the vorticity triplet $(R,S,\omega_{z})$ along the line connecting the centers of the thermal vortices.}
	\label{compare_abc}
\end{figure}
\begin{figure}[htbp]
	\centering
	\subfigure[Hydrodynamic regime, $(\beta_{R},\beta_{N})=(0,100)$]{
		\begin{minipage}[t]{1.0\linewidth}
			\centering
			\includegraphics[width=1.0\columnwidth,trim={0cm 5cm 0cm 4cm},clip]{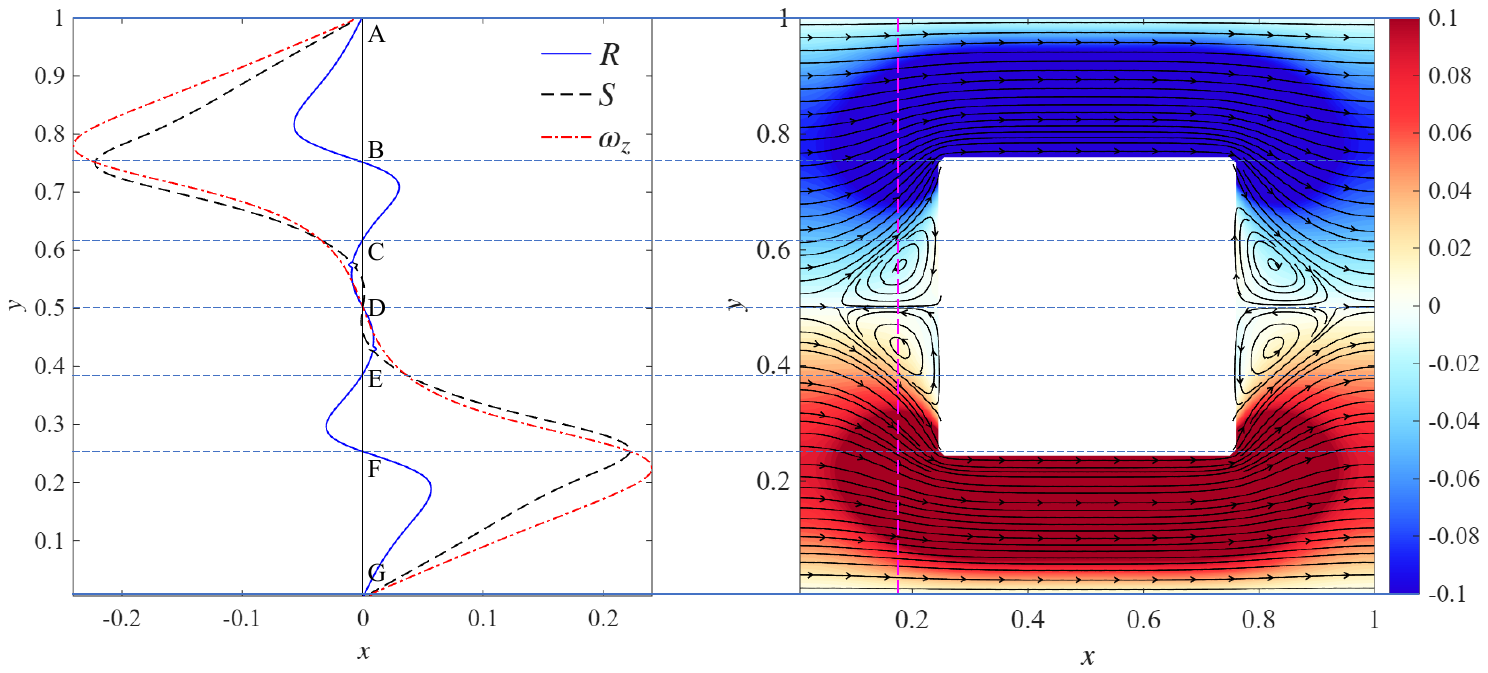}
			\label{compare_Hydrodynamic_Kn001}
		\end{minipage}%
	}%
	
	\subfigure[Hydrodynamic regime, $(\beta_{R},\beta_{N})=(0,10)$]{
		\begin{minipage}[t]{1.0\linewidth}
			\centering
			\includegraphics[width=1.0\columnwidth,trim={0cm 5cm 0cm 4cm},clip]{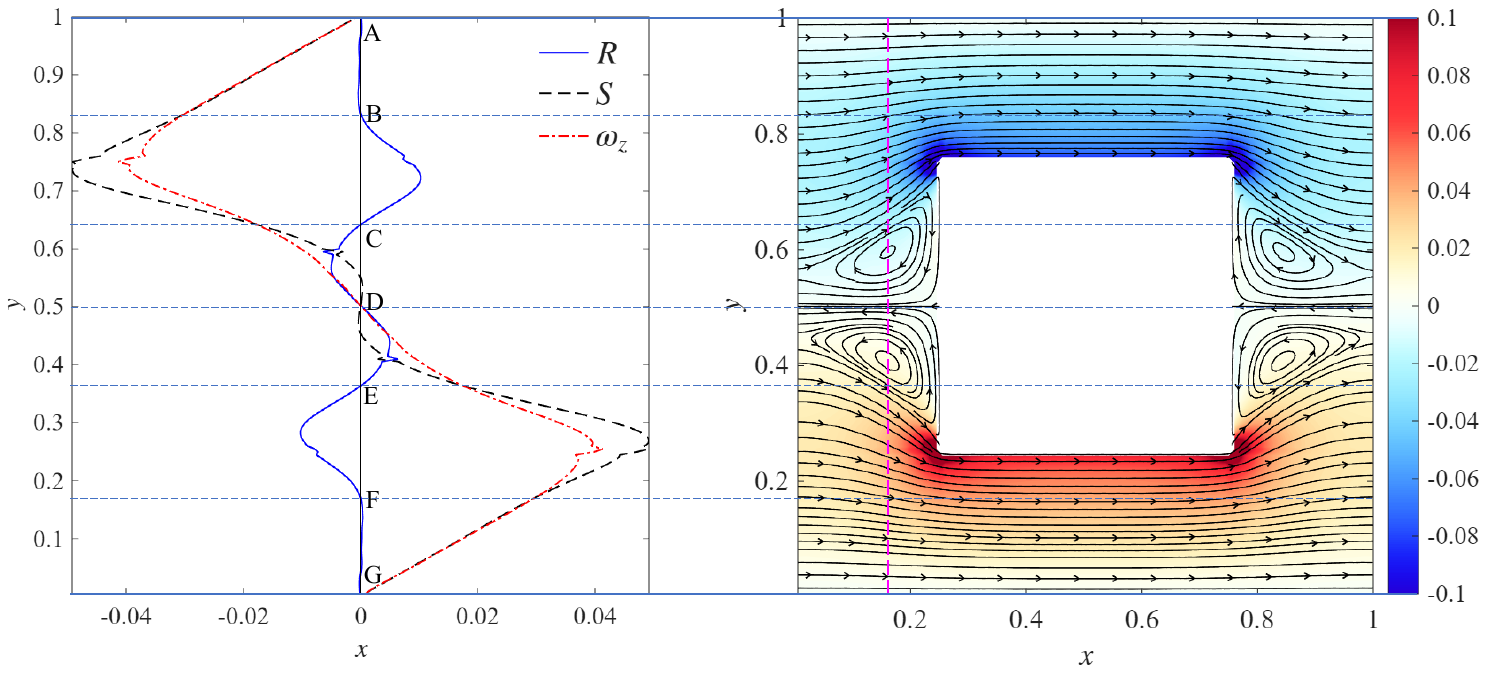}
			\label{compare_Hydrodynamic_Kn01}
		\end{minipage}%
	}%
	\caption{Continuation of Fig. 8. Comparison of the spatial distributions of the vorticity triplet $(R,S,\omega_{z})$ along the line connecting the centers of the thermal vortices.}
	\label{compare_abc1}
\end{figure}

\begin{figure}[htbp]
	\centering
	\subfigure[$R$]{
		\begin{minipage}[t]{0.33\linewidth}
			\centering
			\includegraphics[width=1.0\columnwidth,trim={3cm 0cm 3cm 0cm},clip]{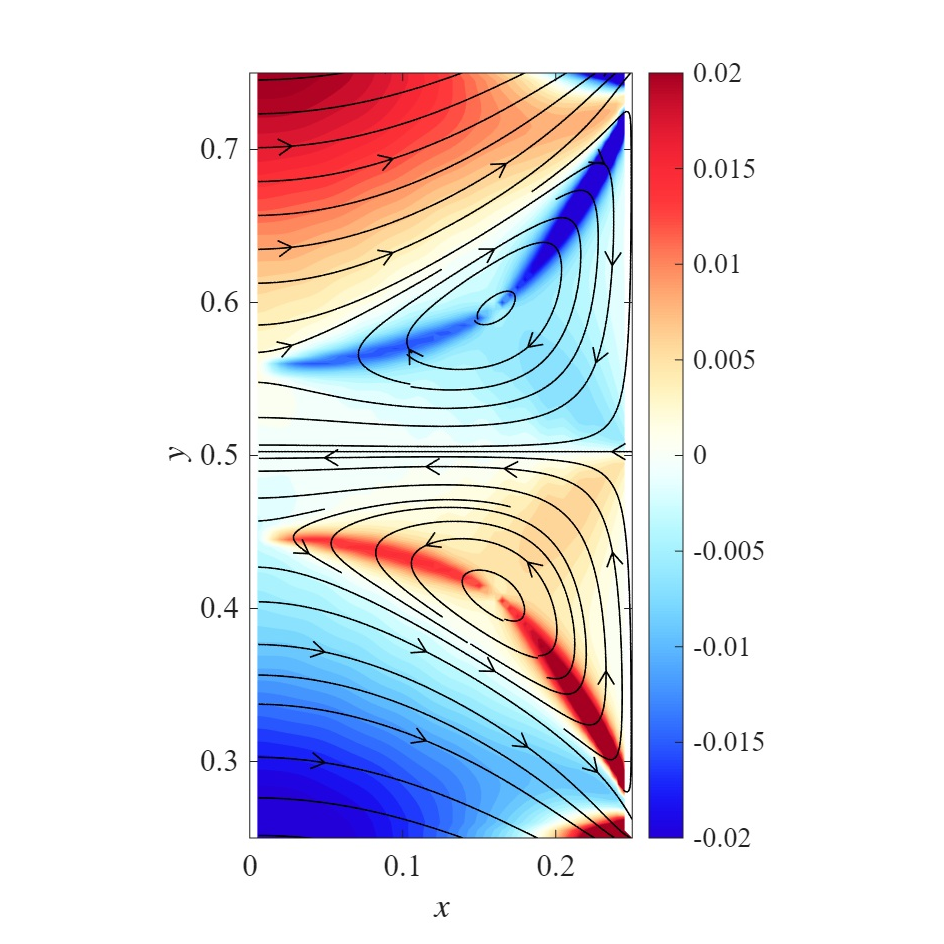}
			\label{R_zoomin}
		\end{minipage}%
	}%
	\subfigure[$S$]{
		\begin{minipage}[t]{0.33\linewidth}
			\centering
			\includegraphics[width=1.0\columnwidth,trim={3cm 0cm 3cm 0cm},clip]{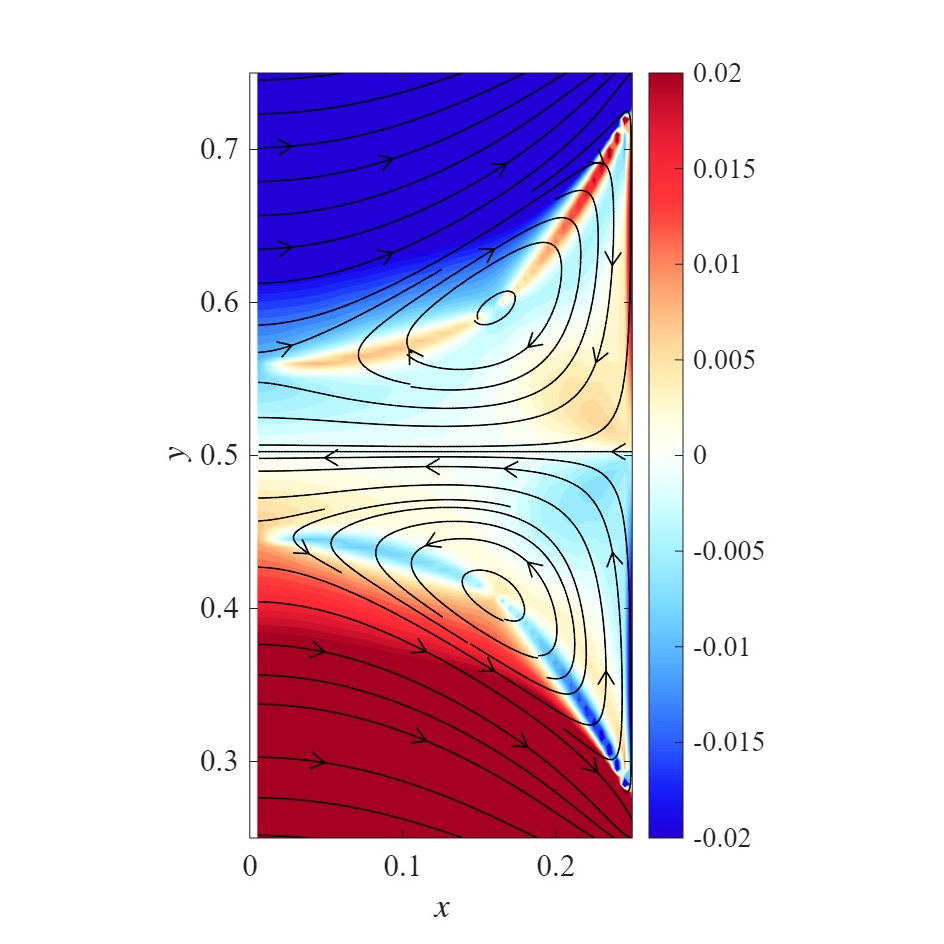}
			\label{S_zoomin}
		\end{minipage}%
	}%
		\subfigure[$\omega_{z}$]{
		\begin{minipage}[t]{0.33\linewidth}
			\centering
			\includegraphics[width=1.0\columnwidth,trim={3cm 0cm 3cm 0cm},clip]{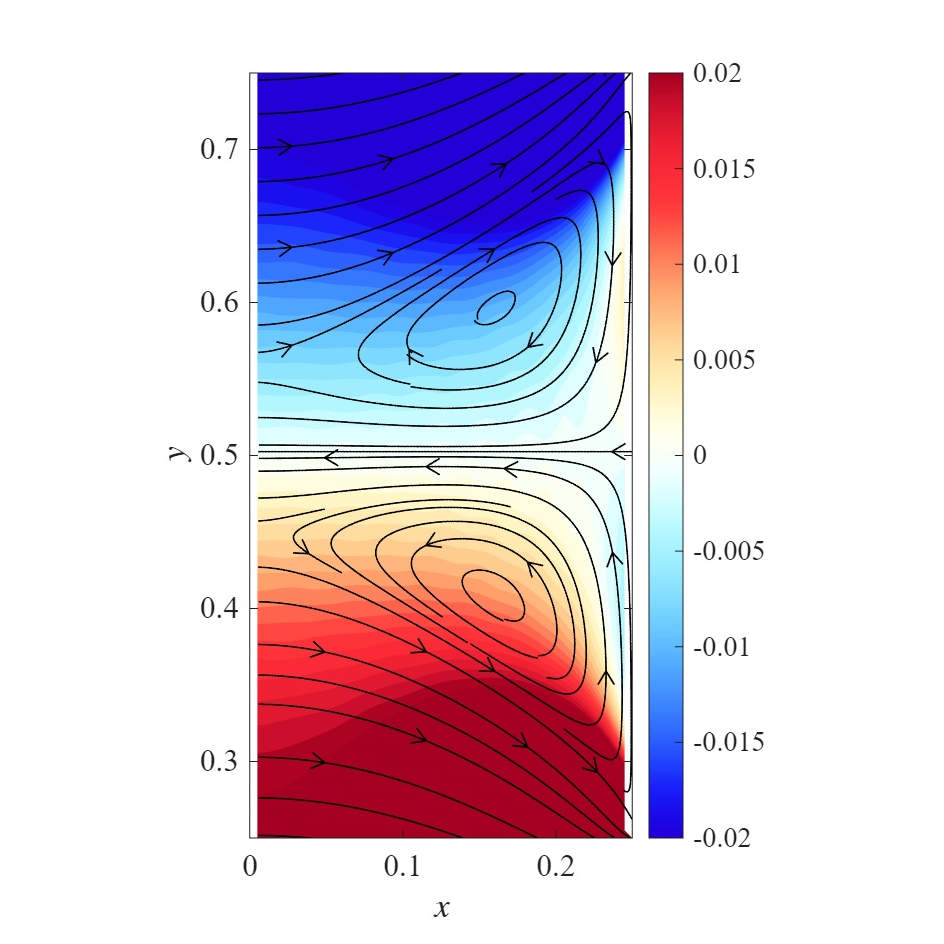}
			\label{OM_zoomin}
		\end{minipage}%
	}%
	\caption{Zoom-in plots of the vorticity triplet $(R,S,\omega_{z})$ within the heat vortices for the hydrodynamic regime $(\beta_{R},\beta_{N})=(0,10)$.}
	\label{zoom_in_views_hydro}
\end{figure}
In contrast, as shown in figure~\ref{T_diffusive}, no heat vortices exist in the diffusive regime where $(\beta_{R},\beta_{N})=(100,0)$ or $(\tau_{R},\tau_{N})=(0.01,\infty)$. In fact, in the diffusive limit $(\beta_{R},\beta_{N})=(\infty,0)$, the phonon transport is dominated by the $R$ scattering process, which causes the heat-flux dissipation. At the steady state, performing the Chapman-Enskog analysis of~\eqref{BTE2} yields $e= e_{R}^{eq}-\tau_{R}\bm{v}\bm{\cdot}\bm{\nabla}e_{R}^{eq}+\mathcal{O}(\tau_{R}^2)$, from which the classical Fourier's law is recovered at the order $\mathcal{O}(\tau_{R})$, i.e., $\bm{q}=\bm{q}_{c}+\mathcal{O}(\tau_{R}^2)$, where $\kappa_{0}=Cv_{g}^2\tau_{R}/2$ is the thermal conductivity coefficient, and the Fourier's law is formulated as
\begin{equation}
	\bm{q}_{c}=-\kappa_{0}\bm{\nabla}T.
\end{equation}
Therefore, we obtain the null vorticity $(\bm{\omega}=\bm{0})$ at the order $\mathcal{O}(\tau_{R})$ due to $\bm{\nabla}\times\bm{\nabla}T=\bm{0}$, as shown in figure~\ref{Vorticity_diffusive}. This result arises from the mutual cancellation of the orbital and spin components, i.e., $R=-S$, as demonstrated in figures~\ref{diffusive},~\ref{Smode_diffusive}, and~\ref{compare_Diffuse}. Small visible errors are concentrated in the vicinity of the pore corners, which could be attributed to the finite $\beta_{R}$ effect.
When $\beta_{R}\rightarrow0$ (or $\tau_{R}\rightarrow\infty$) and $\beta_{N}\gg\beta_{R}$, the phonon transport dominated by the frequent $N$ scattering falls into the hydrodynamic regime, as shown in figures~\ref{T_hydroKn0p01} $(\beta_{N}=100)$ and~\ref{T_hydroKn0p1} $(\beta_{N}=10)$. In this limit,~\eqref{eq29b} also reduces to $\partial_{t}\bm{q}+\bm{\nabla}\bm{\cdot}\bm{Q}=\bm{0}$ owing to the absence of the $R$ scattering. Unlike in the ballistic regime, the heat-flux current tensor $\bm{Q}$ depends on the $N$ scattering process. It is observed that the size of vortices becomes smaller as $\beta_{N}$ increases due to weaker viscous diffusion.
In figures~\ref{Vorticity_hydroKn0p01} and~\ref{Vorticity_hydroKn0p1}, two large block regions with opposite-sign vorticity component are identified in the upper and lower half domains, respectively. In contrast, the vorticity magnitude within the heat vortices is much weaker. As shown in figures~\ref{Rmode_hydroKn0p01} and~\ref{Rmode_hydroKn0p1}, the orbital rotation $R$ enhances the vorticity magnitude near the square pore corners while reducing it in the block regions adjacent to the left and right sides of the unit square cell. The spin $S$ is the primary contributor to the vorticity pattern in the bulk region outside the heat vortices (figures~\ref{Smode_hydroKn0p01} and~\ref{Smode_hydroKn0p1}), whereas the orbital-rotation $R$ dominates the vorticity pattern within the majority of the heat vortices. Figure~\ref{zoom_in_views_hydro} shows the zoom-in views of the vorticity triplet corresponding to the hydrodynamic regime $(\beta_{R},\beta_{N})=(0,10)$. In a small neighborhood of the pore boundary, the vorticity is dominated by $S$, which resembles the boundary layer for fluid flows. Moreover, figures~\ref{compare_Hydrodynamic_Kn001} and~\ref{compare_Hydrodynamic_Kn01} show that along the straight line connecting the heat vortices, the vorticity magnitude is enhanced due to the synergistic effect induced by $R$ and $S$ with the same signs. As $\beta_{R}$ increases from $0$ to $1$, the heat vortices disappear in the transition regimes for both $\beta_{N}=100$ (figure~\ref{T_U1N100}) and $\beta_{N}=10$ (figure~\ref{T_U1N10}), although the overall patterns of the vorticity triplet are similar to the hydrodynamic regimes (see figures~\ref{omegaz_U1N100},~\ref{omegaz_U1N10},~\ref{Rmode_U1N100},~\ref{Rmode_U1N10},~\ref{Smode_U1N100},~\ref{Smode_U1N10}).

The physical mechanism underlying boundary heat-vorticity generation is analyzed in the hydrodynamic limit, where the relaxation time of the $N$ scattering process is much shorter than the $R$ scattering process, i.e., $\beta_{N}\gg 1\gg\beta_{R}$. One can introduce a small dimensionless parameter $\varepsilon\equiv\tau_{N}/\tau_{R}\ll 1$, allowing the phonon distribution function to be expanded as $f=f_{0}+\varepsilon f_{1}+\varepsilon^2 f_{2}+\cdots$~\citep{Banach2008chapman}. By applying the Chapman-Enskog (C-E) multiscale expansion, we arrive at the Guyer-Krumhansl (G-K) equation~\citep{Guyer1966t,Shang2020HeatVortex}
\begin{equation}\label{GK_eq1}
	\frac{\partial\bm{q}}{\partial t}+\frac{\kappa_0}{\tau_{R}}\bm{\nabla}T+\frac{\bm{q}}{\tau_{R}}=\eta\left(\nabla^2\bm{q}+2\bm{\nabla}\vartheta\right)-\zeta\bm{\nabla}\vartheta,
\end{equation}
where $\eta=v_{g}^{2}\tau_{N}/4$ and $\zeta=v_{g}^{2}\tau_{N}/2$ are the first and second viscosity coefficients (for 2D materials), respectively. By introducing the potential function $\phi\equiv\kappa_{0}T/\tau_{R}+(\zeta-2\eta)\vartheta$, a more compact form of~\eqref{GK_eq1} can be written as
	\begin{equation}\label{GK_eq2}
	\frac{\partial\bm{q}}{\partial t}+\bm{\nabla}\phi+\frac{\bm{q}}{\tau_{R}}=\eta\nabla^2\bm{q}.
	\end{equation}
The heat-flux diffusion term $\eta\nabla^2\bm{q}$ is analogous to the momentum diffusion term $\nu\nabla^2\bm{u}$ in the NS equation ($\nu$ is the kinematic viscosity), and $\phi$ resembles the hydrodynamic pressure. The nonlinear convection term $\bm{u}\bm{\cdot}\bm{\nabla}\bm{u}$ does not appear here and is instead replaced by a linear term $\bm{q}/\tau_{R}=(\alpha/\tau_R)\bm{u}$. Since $\bm{\nabla}\times\bm{\nabla}\phi=\bm{0}$ and $\bm{\omega}\equiv\bm{\nabla}\times\bm{q}$, taking the curl of~\eqref{GK_eq2} yields the transportation equation for the heat vorticity:
\begin{equation}\label{vorticity_eq}
	\frac{\partial\bm{\omega}}{\partial t}+\frac{\bm{\omega}}{\tau_{R}}=\eta\nabla^2\bm{\omega}.
\end{equation}
Compared with the Helmholtz vorticity equation for 2D fluid flows, the nonlinear convection term $\bm{u}\bm{\cdot}\bm{\nabla}\bm{\omega}$ is replaced by the linear term $\bm{\omega}/\tau_{R}$.

Consider a fixed control domain $\mathscr{V}$ enclosed by the boundary surface (curve) $\partial\mathscr{V}$ oriented by the outward unit normal vector $\hat{\bm{n}}$. The total vorticity contained within $\mathscr{V}$ is given by
\begin{equation}\label{Gamma_omega}
\bm{\Gamma}\equiv\int_{\mathscr{V}}\bm{\omega}dV.
\end{equation}
By using the Gauss-Ostrogradsky divergence theorem, the volume integral of the vorticity diffusion term is transformed into the surface integral as
\begin{equation}\label{xx1}
\int_{\mathscr{V}}\eta\nabla^2\bm{\omega}dV=\oint_{\partial\mathscr{V}}\eta\hat{\bm{n}}\bm{\cdot}\bm{\nabla}\bm{\omega}dS=-\oint_{\partial\mathscr{V}}\bm{\sigma}dS,
\end{equation}
where the unit normal $\bm{n}=-\hat{\bm{n}}$ points toward the interior of the integral domain. The integrand $\bm{\sigma}$ can be referred to as the boundary heat vorticity flux (BHVF), which is defined as
\begin{equation}\label{xx2}
\bm{\sigma}\equiv\eta\bm{n}\bm{\cdot}\bm{\nabla}\bm{\omega}=\eta\frac{\partial\bm{\omega}}{\partial n}\quad~\text{on}~\partial\mathscr{V}.
\end{equation}
The BHVF can be interpreted as an analogy to the boundary vorticity flux (BVF) $\bm{\sigma}_{f}\equiv\nu\partial_{n}\bm{\omega}$ in fluid dynamics (where $\nu$ is the kinematic viscosity), which was first introduced by~\citet{Lighthill1963,Lighthill1986} to explain the boundary vorticity generation by the surface pressure gradient for incompressible viscous flow interacting with a no-slip wall.
Physically, since no heat vorticity exists outside the region occupied by the material, the BHVF quantifies the boundary production rate of the heat vorticity for solid materials and thus represents the strength of the heat vorticity source distributed on a boundary. 

\begin{figure}[t]
	\centering
	\subfigure[]{
		\begin{minipage}[t]{0.5\linewidth}
			\centering
			\includegraphics[width=0.97\columnwidth,trim={0cm 0cm 0.0cm 0cm},clip]{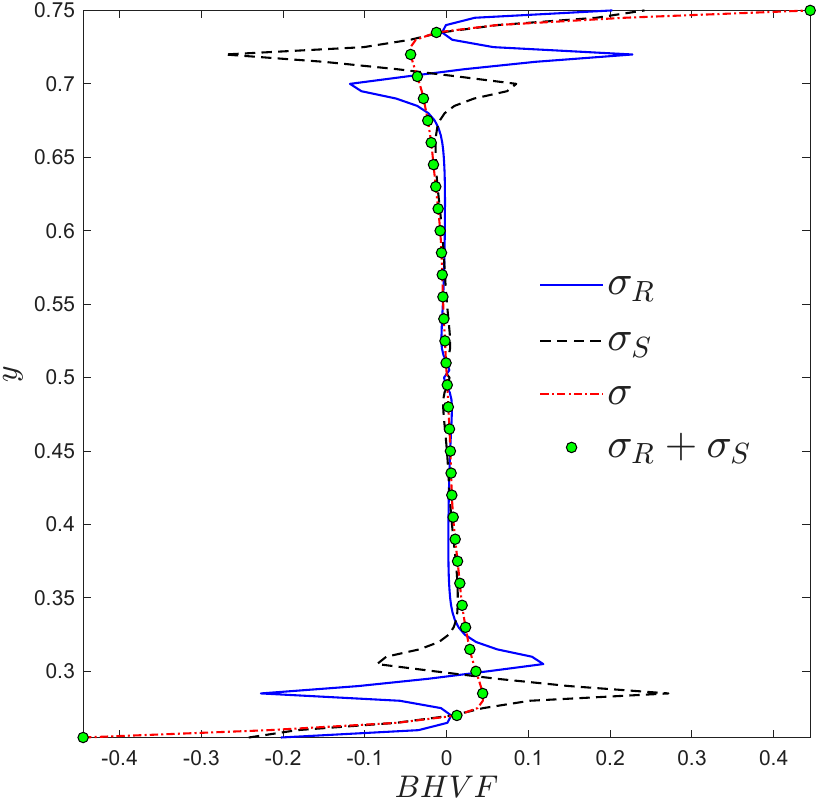}
			\label{BHVF1}
		\end{minipage}%
	}%
	\subfigure[]{
		\begin{minipage}[t]{0.5\linewidth}
			\centering
			\includegraphics[width=1.0\columnwidth,trim={0cm 0cm 0.0cm 0cm},clip]{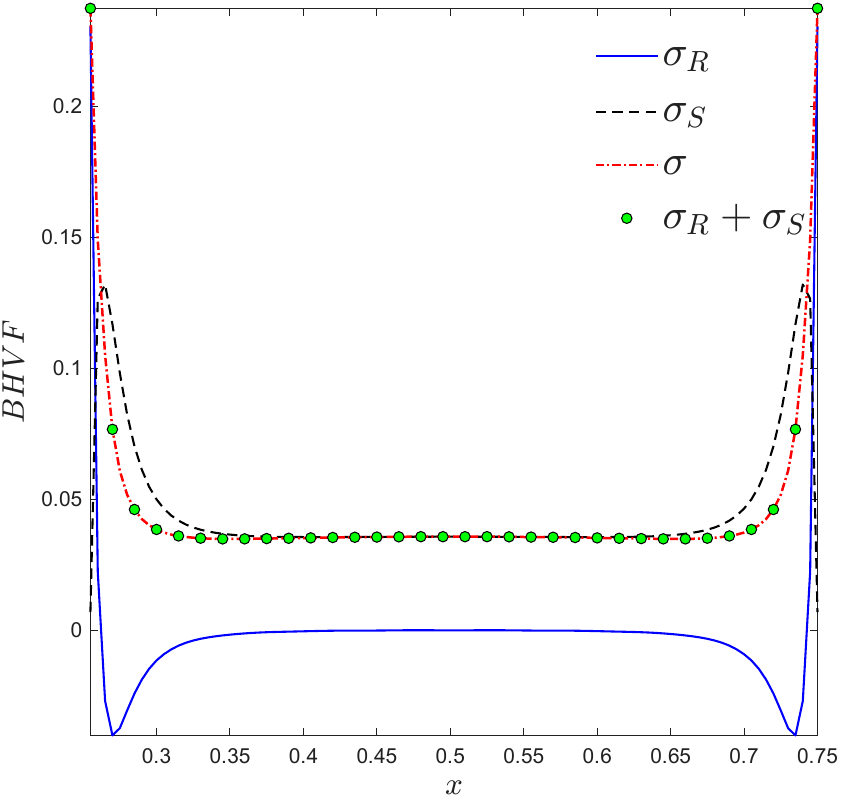}
			\label{BHVF2}
		\end{minipage}%
	}%
	\caption{Distributions of the BHVF triplet $(\sigma_{R},\sigma_{S},\sigma)$ along (a) the front pore boundary $(x=0.25,y\in[0.25,0.75])$ and (b) the upper pore boundary $(x\in[0.25,0.75],y=0.75)$ in the hydrodynamic regime $(\beta_{R},\beta_{N})=(0,10)$.}
	\label{BHVF12}
\end{figure}

By virtue of the solenoidal condition  $\bm{\nabla}\bm{\cdot}\bm{\omega}=\bm{0}$,~\eqref{xx2} can be decomposed as
\begin{equation}\label{eq5p7}
\bm{\sigma}=\eta\left(\bm{n}\times\bm{\nabla}\right)\times\bm{\omega}-\eta\bm{n}\times(\bm{\nabla}\times\bm{\omega}).
\end{equation}
The first term on the right-hand side of~\eqref{eq5p7} is evaluated as
\begin{equation}\label{eq5p8}
\eta\left(\bm{n}\times\bm{\nabla}\right)\times\bm{\omega}=\eta\bm{\nabla}_{\pi}\omega_{n}+\eta\bm{K}\bm{\cdot}\bm{\omega}_{\pi}-\eta(\bm{\nabla}_{\pi}\bm{\cdot}\bm{\omega})\bm{n}.
\end{equation}
In~\eqref{eq5p8}, the heat vorticity is decomposed as $\bm{\omega}=\bm{\omega}_{\pi}+\omega_{n}\bm{n}$, where $\bm{\omega}_{\pi}$ and $\omega_{n}$ denote the tangential and boundary-normal heat vorticity components, respectively. $\bm{\nabla}_{\pi}\equiv\bm{\nabla}-\bm{n}\bm{n}\bm{\cdot}\bm{\nabla}$ represents the surface tangential gradient operator. The surface curvature tensor is given by $\bm{K}\equiv-\bm{\nabla}_{\pi}\bm{n}$, which vanishes for a flat boundary. Notably, the terms in~\eqref{eq5p8} can be nonzero only in 3D cases. Upon applying~\eqref{vorticity_eq} on $\partial\mathscr{V}$, the second term on the right-hand side of~\eqref{eq5p7} evaluates to 
\begin{equation}\label{eq5p9}
-\eta\bm{n}\times(\bm{\nabla}\times\bm{\omega})=\bm{n}\times\frac{\partial\bm{q}}{\partial t}+\bm{n}\times\bm{\nabla}_{\pi}\left(\phi-\eta\vartheta\right)+\bm{n}\times\frac{\bm q}{\tau_{R}}.
\end{equation}
Substituting~\eqref{eq5p8} and~\eqref{eq5p9} into~\eqref{eq5p7} yields the general expression for the BHVF:
\begin{eqnarray}\label{eq5p10}
\bm{\sigma}&=&\bm{n}\times\frac{\partial\bm{q}}{\partial t}+\bm{n}\times\bm{\nabla}_{\pi}\left(\phi-\eta\vartheta\right)+\bm{n}\times\frac{\bm q}{\tau_{R}}\nonumber\\
& &+\eta\bm{\nabla}_{\pi}\omega_{n}+\eta\bm{K}\bm{\cdot}\bm{\omega}_{\pi}-\eta(\bm{\nabla}_{\pi}\bm{\cdot}\bm{\omega})\bm{n}.
\end{eqnarray}
Equation~\eqref{eq5p10} unveils all physical mechanisms for the generation of heat vorticity on a boundary in the hydrodynamic limit.
For the 2D phonon transport sketched in figure~\ref{porous_material},~\eqref{eq29a} implies that $\vartheta=\bm{\nabla}\bm{\cdot}\bm{q}=0$ at the steady state and that $\phi=\kappa_{0}T/\tau_{R}=\frac{1}{2}Cv_{g}^2 T$. For the flat boundary of the square pore, $\bm{K}=\bm{O}$ holds. Consequently, at the steady state,~\eqref{eq5p10} simplifies
\begin{eqnarray}\label{eq5p11}
	\bm{\sigma}=\bm{n}\times\bm{\nabla}_{\pi}\phi+\bm{n}\times\frac{\bm q}{\tau_{R}}=\bm{n}\times\frac{\bm{q}-\bm{q}_{c}}{\tau_{R}}.
\end{eqnarray}
Equation~\eqref{eq5p11} indicates that the boundary generation of heat vorticity is attributed to the tangential component of the deviational heat flux (i.e., $\bm{q}-\bm{q}_{c}$), which represents the heat flux component departing from the classical Fourier's law. The BHVF in~\eqref{xx2} can be decomposed as
\begin{subequations}\label{BHVF_triplet}
	\begin{gather}
		\bm{\sigma}=\bm{\sigma}_{R}+\bm{\sigma}_{S},\label{BHVF_triplet1}\\
		\bm{\sigma}_{R}\equiv\eta\frac{\partial\bm{R}}{\partial n},\qquad
		\bm{\sigma}_{S}\equiv\eta\frac{\partial\bm{S}}{\partial n},\qquad
		\text{on }\partial\mathscr{V}.\label{BHVF_triplet2}
	\end{gather}
\end{subequations}
The $\bm{e}_{z}$-components of the BHVF triplet \((\bm{\sigma}_{R},\bm{\sigma}_{S},\bm{\sigma})\) are denoted by \((\sigma_{R},\sigma_{S},\sigma)\).
Their spatial distributions along the front and upper boundaries of the square pore are shown in figures~\ref{BHVF1} and~\ref{BHVF2}, respectively.  Along the front pore boundary, all BHVFs are concentrated nearby the pore corners. Along the upper pore boundary, ${\sigma}_{R}$ is primarily generated near the pore corners, whereas ${\sigma}_{S}$ is significant at almost all positions and exhibits two high positive peaks near the pore corners. The total BHVF $\sigma$ is dominated by ${\sigma}_{S}$ in most regions.

Owing to the geometrical symmetry about \(y=0.5\), the total vorticity over the entire domain must vanish. To associate the evolution of the vorticity integral with the BHVF, we select the upper half domain $\mathscr{V}_{U}$ as the control volume, bounded by the closed curve $\partial\mathscr{V}_{U}\rightarrow\overline{BAGFPMDCB}$. In general, the total vorticity $\bm{\Gamma}_{U}\equiv\int_{\mathscr{V}_U}\bm{\omega}dV$ is non-zero for $t>0$ and evolves temporally.
From~\eqref{vorticity_eq},~\eqref{Gamma_omega}, and~\eqref{xx1}, we obtain the ordinary differential equation (ODE) governing $\bm{\Gamma}_{U}$: 
\begin{equation}\label{xx3}
\frac{d\bm{\Gamma}_U}{dt}+\frac{\bm{\Gamma}_U}{\tau_{R}}=-\oint_{\partial\mathscr{V}_U}\bm{\sigma}dS\equiv\bm{\Phi}_{U}.
\end{equation}
With the initial condition $\bm{\Gamma}_{U}(t=0)=\bm{0}$, the formal solution to~\eqref{xx3} is 
\begin{eqnarray}\label{Gamma_U_t}
		\bm{\Gamma}_U(t)
		=\int_{0}^{t}
	\bm{\Phi}_{U}(t^\prime)
		\exp\left(-\frac{t-t^\prime}{\tau_R}\right)dt^\prime.
\end{eqnarray}
which indicates that the instantaneous total vorticity in $\mathscr{V}$ is fully determined by the spatiotemporal accumulation of the BHVF, weighted by the memory kernal $\exp\left(-(t-t^\prime)/{\tau_R}\right)$ from the $R$ scattering. As $t\rightarrow\infty$ (steady state), $\bm{\Phi}_{U}(t)\rightarrow\bm{\Phi}_{U}(\infty)$. Then, it follows from~\eqref{eq5p11} and~\eqref{Gamma_U_t} that
\begin{equation}\label{Gamma_U_infty}
\bm{\Gamma}_{U}(\infty)=\tau_{R}\bm{\Phi}_{U}(\infty)=\oint_{\partial\mathscr{V}_U}\widehat{\bm{n}}\times\left(\bm{q}-\bm{q}_{c}\right)dS.
\end{equation}
It is noted that the steady value in~\eqref{Gamma_U_infty} can also be derived directly by setting $d/dt=0$ in~\eqref{xx3}. A similar analysis can be carried out for the lower half domain $\mathscr{V}_{L}$ enclosed by the boundary curve $\partial\mathscr{V}_{L}\rightarrow\overline{AOEDMNHGA}$. In analogy to~\eqref{Gamma_U_infty}, for $\bm{\Gamma}_{L}\equiv\int_{\mathscr{V}_L}\bm{\omega}dV$, we have $\bm{\Phi}_{L}(t)\rightarrow\bm{\Phi}_{L}(\infty)$ at the steady state. Then, it follows that
\begin{equation}\label{Gamma_L_infty}
	\bm{\Gamma}_{L}(\infty)=\tau_{R}\bm{\Phi}_{L}(\infty)=\oint_{\partial\mathscr{V}_L}\widehat{\bm{n}}\times\left(\bm{q}-\bm{q}_{c}\right)dS.
\end{equation}
Summing~\eqref{Gamma_U_infty} and~\eqref{Gamma_L_infty} recovers the zero total vorticity in the control domain $\mathscr{V}$:
\begin{eqnarray}
\bm{\Gamma}&=&\bm{\Gamma}_{U}+\bm{\Gamma}_{L}={\tau}_{R}\bm{\Phi}(\infty)\nonumber\\
&=&\oint_{\partial\mathscr{V}}\widehat{\bm{n}}\times\left(\bm{q}-\bm{q}_{c}\right)dS\nonumber\\
&=&\oint_{\partial\mathscr{V}}\widehat{\bm{n}}\times\bm{q}dS-\int_{\partial\mathscr{V}}\bm{\nabla}\times\bm{q}_{c}dS\nonumber\\
&=&\oint_{\partial\mathscr{V}}\widehat{\bm{n}}\times\bm{q}dS\nonumber\\
&=&\bm{0}.
\end{eqnarray}
Note that when the slip effect of the heat flux along the boundary is significant, an additional heat-vorticity layer should be included in the expression for the total vorticity. However, this is not the primary concern of the present study.

\section{Conclusions and Discussions}\label{Conclusions and Outlook}
\begin{enumerate}
	\item From a fluid-mechanical perspective, we propose a direction-dependent decomposition of the heat vorticity for kinetic phonon transport in 2D solid materials. The heat vorticity $\bm{\omega}\equiv\bm{\nabla}\times\bm{q}$, being proportional to the curl of the phonon drift velocity, is decomposed into the orbital-rotation component $\bm{R}$ and the spin component $\bm{S}$~\eqref{vd}. Together, these three quantities constitute the heat-vorticity triplet $(\bm{R},\bm{S},\bm{\omega})$. Compared with the conventional single heat-vorticity paradigm, the resulting framework extracts more detailed vortical-flow information from the heat-flux field for different kinetic regimes.
	\item Heat conduction in a periodic 2D porous solid material with aligned square pores is simulated based on the Peierls-Boltzmann equation under the Callway dual-relaxation-time approximation. In the ballistic regime, the vorticity is dominated by $\bm{S}$, with a negligible contribution from $\bm{R}$. The profiles of $\bm{S}$ and $\bm{\omega}$ exhibit high peaks within the streaky structures between twin heat vortices and the outer regions. In the diffusive regime, the heat vortices disappear due to rare phonon scattering, and the vanishing Fourier vorticity is interpreted as a cancellation between $\bm{R}$ and $\bm{S}$. In the hydrodynamic regime, the vortices become smaller due to more frequent N scattering and resulting weaker viscous diffusion effect. It is observed that the vorticity pattern is primarily dominated by $\bm{S}$ outside the heat vortices and by $\bm{R}$ within the majority of the heat vortices. Near the pore boundary, the vorticity is dominated by $\bm{S}$, resembling the boundary-layer flow in viscous fluids. In addition, the heat vortices disappear in the transition regimes, while the overall patterns of the vorticity triplet resemble those in the hydrodynamic regime. In summary, the emergence of heat vortices and the vorticity triplet patterns can serve as effective macroscopic signatures of distinct kinetic regimes.
	\item The boundary generation mechanism of heat vorticity is discussed based on the Guyer-Krumhansl (G-K) equation in the hydrodynamic limit. By analogy with Lighthill's boundary vorticity flux (BVF) for fluids, we introduce a new physical concept, the boundary heat vorticity flux (BHVF) $\bm{\sigma}$~\eqref{xx2}, to describe the boundary sources of heat vorticity in solid materials, where all boundary generation mechanisms of heat vorticity are clearly uncovered by~\eqref{eq5p10} for general 2D or 3D materials. In particular, for 2D steady hydrodynamic phonon transport, it is found that the BHVF is solely attributed to the tangential component of the net heat flux deviating from the classical Fourier heat flux~\eqref{eq5p11}. The total vorticity contained in a control volume is shown to arise from the spatiotemporal accumulation of the BHVF, weighted by the memory kernel from the R scattering (\eqref{xx3} and \eqref{Gamma_U_t}). The vorticity triplet naturally gives rise to the BHVF triplet $(\bm{\sigma}_{R},\bm{\sigma}_{S},\bm{\sigma})$, which characterizes the boundary sources of orbital rotation and spin~\eqref{BHVF_triplet}.
	\item In future work, we intend to study kinetic phonon transport in 2D porous solid material with circular holes by solving the full Boltzmann equation without invoking linearization assumption, for example, like the model of~\citet{Sykora2023}. The vorticity decomposition approach will be employed to analyze heat vortex structures such as the von K\'{a}rm\'{a}n vortex street of the heat-flux field.
\end{enumerate}

\section*{Declarations}
\begin{itemize}
	\item \textbf{Funding} 	This work was funded by the National Natural Science
	Foundation of China (Grant No. 12402262 and 52506078) and the Zhejiang Provincial Natural Science Foundation of China under Grant No. LMS26E060012.
	\item \textbf{Conflict of interest} The authors have no conflicts to disclose.
	\item \textbf{Data availability} No datasets is generated for the present study.
	\item \textbf{Author contributions}	Tao Chen and Chuang Zhang are involved in conceptualization, formal analysis, funding acquisition, investigation, visualization, methodology, the original draft preparation, review and editing of the manuscript. Tianshu Liu is involved in the review and editing of the manuscript.
	\item \textbf{Author ORCID} Tao Chen https://orcid.org/0000-0001-6838-204X; Chuang Zhang  https://orcid.org/0000-0002-2136-0799; Tianshu Liu https://orcid.org/0000-0001-6297-1660.
\end{itemize}

\backmatter


\begin{appendices}

\end{appendices}


\bibliography{sn-bibliography}

\end{document}